\documentclass{article}

\usepackage[english]{babel}
\usepackage[a4paper,top=2cm,bottom=2cm,left=2.5cm,right=2.5cm]{geometry}
\usepackage{amsmath,amssymb}
\usepackage{booktabs}
\usepackage{caption}
\usepackage{graphicx}
\usepackage{microtype}
\usepackage[section]{placeins}  % keep floats inside their section
\usepackage[skip=\baselineskip]{parskip}
\usepackage[colorlinks=true,allcolors=blue]{hyperref}
\usepackage{authblk}

\title{Advantage-level Aggregation Reinforcement Learning for X-point Target Magnetic Configuration Control in an EXL-50U Experiment-Calibrated Simulation Environment}

\author[1,2,3]{Siqi Ding}
\author[1,2,3]{Xuanhe Wang}
\author[4]{Pei Guo}
\author[1,2,3]{Guoyang Shi}
\author[5]{Changquan Yu}
\author[6]{Yiting Wang}
\author[1,2,3]{Xianming Song}
\author[1,2,3]{Xiang Gu}
\author[1,2,3]{Zhengyuan Chen}
\author[1,2,3]{Lei Xing}
\author[1,2,3]{Yapeng Zhang}
\author[1,2,3]{Jianguo Chen}
\author[1,2,3,*]{Tianyuan Liu}

\affil[1]{Beijing ENN Fusion Energy Science and Technology Co., Ltd., Beijing 101111, China}
\affil[2]{Beijing Key Laboratory of High Magnetic Field Spherical Torus Fusion Energy, Beijing 101111, China}
\affil[3]{Hebei Key Laboratory of Compact Fusion, Langfang 065000, China}
\affil[4]{Key Laboratory of Materials Modification by Beams of the Ministry of Education, School of Physics, Dalian University of Technology, Dalian 116024, China}
\affil[5]{School of Nuclear Science and Engineering, East China University of Technology, Nanchang 330013, Jiangxi, China}
\affil[6]{School of Mechanics and Engineering Science, Peking University, Beijing 100084, China}
\date{}

\begin{document}
\maketitle
\vspace{-3em}% 按观感调到 -0.8em ~ -2em
\begin{center}
\textit{* Corresponding authors: Tianyuan Liu (liutianyuan@enn.cn).}
\end{center}

\begin{abstract}
Managing divertor heat loads is a central challenge for compact, high-power
tokamaks.  To increase local flux expansion and create a dissipation volume
magnetically decoupled from the core, the compact spherical-torus design EHL-2
adopts the X-point target (XPT) as its reference divertor configuration.
Realising these benefits requires the secondary X-point to remain on the
divertor leg; its displacement can degrade the intended topology and exhaust
geometry.  Existing experiments, including recent EXL-50U discharges, still
rely mainly on precomputed feedforward waveforms with
proportional--integral--derivative (PID) loops on global quantities.  The
secondary null is not yet under dedicated closed-loop feedback, so XPT
operation remains repeatable but not yet routine.  We formulate reconstructed-null XPT
feedback as a multi-objective reinforcement learning (RL) control problem in a free-boundary environment
calibrated to EXL-50U discharge \#13906.  To address the strong coupling among plasma current, boundary shape, and
multiple null constraints---where conventional reward-level scalarisation
collapses objective-specific temporal credit into a single advantage---we
develop Advantage Aggregation (AdvA), which preserves objective-wise
temporal credit assignment before worst-objective-aware nonlinear
scalarisation and introduces a controlled residual correction to the policy
update.
Instantiated with proximal policy optimisation (PPO), AdvA-PPO is evaluated
against conventional reward-level PPO and the
experiment-derived feedforward-plus-PID baseline under nominal operation,
measurement uncertainties, and unseen initial equilibria.  On the nominal
$500\,\mathrm{ms}$ rollout, AdvA-PPO raises the mean worst-channel score from
$0.23$ to $0.81$ relative to Reward-PPO and reduces mean X-point flux
root-mean-square error (RMSE) by
about $20\times$.  Under combined measurement uncertainties, it is the only
learned controller that completes the horizon while retaining a usable XPT
shape.  Multi-initialization fine-tuning further enables a single AdvA-PPO
policy to complete full-horizon operation across both divertor and limiter
initial equilibria.  These results provide a simulation-based foundation for
future real-time XPT validation on EXL-50U.

\end{abstract}


% -----------------------------------------------------------------------------
\section{Introduction}
\label{sec:intro}

Magnetic-confinement fusion offers a route to abundant, low-carbon energy, and
the tokamak is among its most developed concepts.  A reactor-scale tokamak,
however, must exhaust intense particle and power fluxes without exceeding the
steady-state limits of plasma-facing materials.  Extrapolations for compact,
high-power devices illustrate the severity of this challenge
\cite{kuang2020sparc}.  Advanced divertor configurations (ADCs) address it by
reshaping the scrape-off-layer magnetic geometry to increase connection
length, flux expansion, wetted area, and the volume available for power and
momentum dissipation.  Their exhaust benefit therefore depends not only on
edge-plasma physics, but also on the ability to establish and maintain the
intended magnetic topology.

Experiments have realized several ADC families, including the snowflake (SF),
quasi-snowflake (QSF), Super-X, X-divertor (XD), and X-point target (XPT).
Their exhaust benefits are well established: dedicated SF control on DIII-D sustained a $\sim2.5\times$ peak-heat-flux reduction for $2$--$3$\,s,
NSTX SF experiments reduced the peak heat flux between edge-localised modes
(ELMs) by $3$--$5\times$,
and the three-X-point TCV Jellyfish reduced the peak parallel target heat flux
by at least $50\%$ relative to a single null
\cite{kolemen2018diiid,soukhanovskii2016,gorno2024}.  Control maturity,
however, does not follow directly from exhaust performance.  As the number of
nulls increases, plasma current, the main boundary, divertor legs, and strike
points must share a limited set of poloidal-field (PF) actuators, so improving one
quantity can degrade another.

Among these ADCs, the XPT is the configuration most directly relevant to the
EHL-2 divertor route.  It places a secondary magnetic null in the divertor leg,
away from the confined plasma, thereby combining a long leg with strong local
flux expansion and a dissipation volume that is magnetically decoupled from the
core \cite{labombard2015,umansky2017assessment}.  Experiments on TCV and MAST-U
show that this topology improves divertor conditions, sustains an X-point
radiator regime, and passively rejects disturbances near the secondary null
\cite{raj2022,lee2025,lonigro2026,winkel2026}.  These exhaust properties are
conditional on the magnetic topology: the secondary null must remain on the
divertor leg of the primary separatrix to preserve the intended flux expansion
and magnetically decoupled dissipation volume.  Secondary-null displacement can
therefore degrade the functional geometry even when plasma current and global
position remain acceptable.  XPT control is consequently not an ancillary
shape-control task, but an enabling requirement for converting magnetic access
to the configuration into repeatable exhaust capability.

For EHL-2 the XPT is not one option among several but the main reference
divertor configuration: integrated poloidal-field-system and free-boundary
equilibrium design shows that an outer XPT is magnetically accessible within
the coil set, and edge-transport calculations predict the associated heat-load
and detachment behaviour \cite{gu2025ehl2,wang2025ehl2}.  Because the EHL-2
evidence is design and simulation based, it defines both the target topology
and the control requirement that motivate the present study.

\begin{figure}[htbp]
    \centering
    \includegraphics[width=\linewidth]{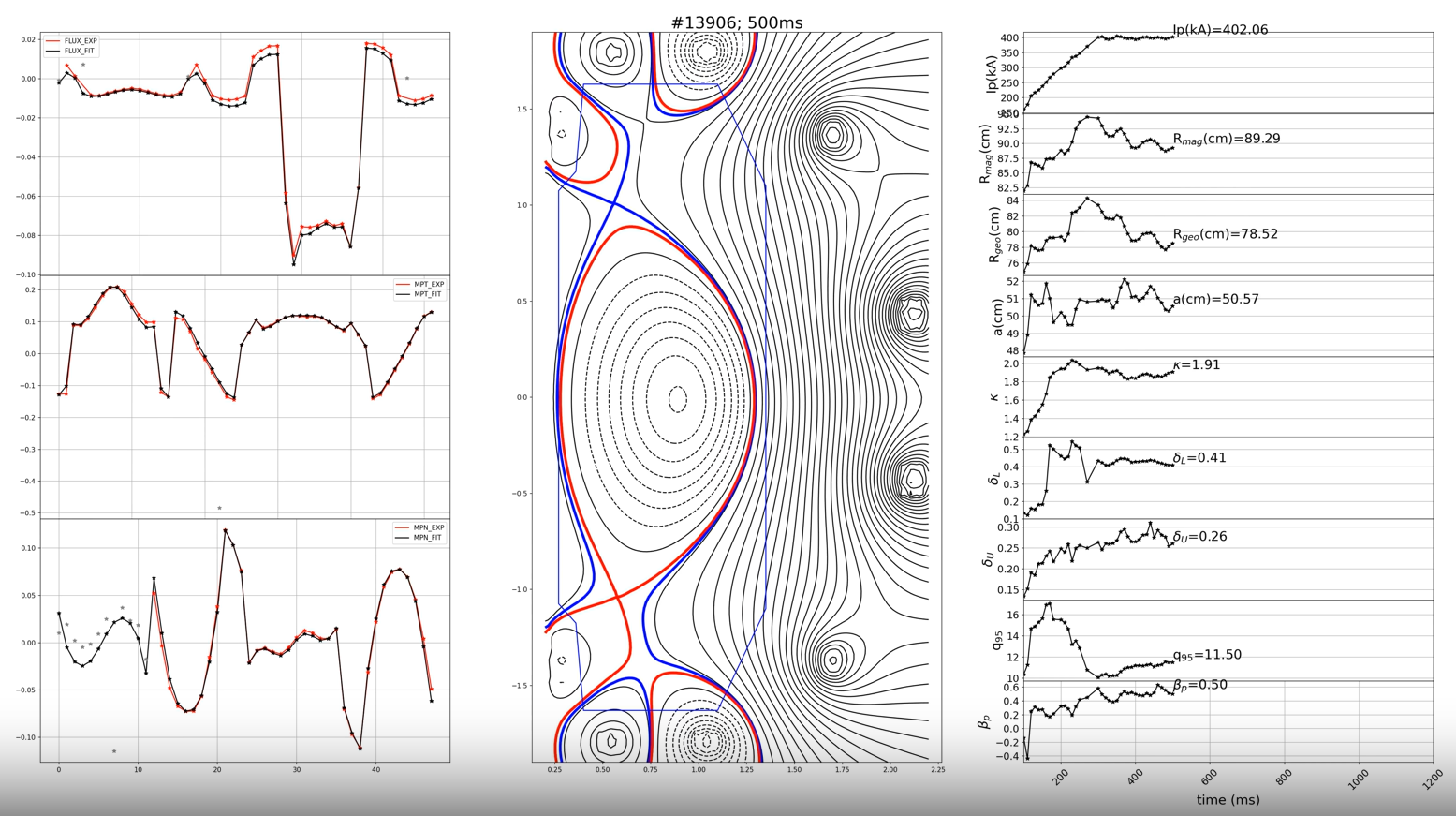}
    \caption{Representative experimental XPT equilibrium in EXL-50U
    discharge \#13906 at $t=500\,\mathrm{ms}$.  The reconstructed magnetic
    topology demonstrates experimental access to XPT operation.}
    \label{fig:exl50u-exp-xpt}
\end{figure}

Motivated by this route, XPT equilibria have now been realized in multiple
EXL-50U discharges.  Figure~\ref{fig:exl50u-exp-xpt} shows a representative
reconstruction from discharge \#13906.  These experiments establish that the
target topology is magnetically accessible on the device, but access in a
reconstructed equilibrium is not equivalent to regulating the
performance-critical secondary null throughout a discharge.  The experiments
still rely primarily on a classical feedforward-plus-PID scheme
(FF+PID): precomputed PF-coil feedforward waveforms
supplemented by PID feedback on global quantities.  Reaching the desired
topology still requires shot-to-shot tuning, and the secondary X-point is not
yet maintained by a dedicated real-time feedback loop.  XPT operation on
EXL-50U should therefore be regarded as repeatably demonstrated but not yet
routine.  A similar commissioning burden was reported on MAST-U, where the
first steady XPT required three successive discharges to tune the feedforward
virtual-circuit waveform \cite{anand2024}.  The EXL-50U experiments provide
physically realized target equilibria and a basis for calibrating the control
environment, rather than evidence of closed-loop secondary-null control.

This limitation is shared by the small number of XPT experiments worldwide,
as summarized in Table~\ref{tab:xpt-control}.  On MAST-U the XPT could only be
obtained transiently in the 2021 campaign under pure feedforward operation;
steady-state XPT became possible once boundary variables were placed under
feedback, but the real-time reconstruction did not estimate the secondary
X-point.  That null was instead created by feedforward virtual-circuit commands
driving the local radial and vertical fields toward zero at a user-specified
location, which left a finite and slowly growing field offset and a
$\sim$10\,cm discrepancy in the achieved null position, and required three
successive discharges to converge \cite{anand2024}.  On TCV, published XPT
campaigns quantify exhaust performance, the X-point radiator regime, and
detachment dynamics under the device's divertor shape control; they do not
report a feedback loop that treats the secondary null as an explicit real-time
control variable together with the main boundary
\cite{theiler2017,raj2022,lee2025,winkel2026}.  Offline reconstruction can
recover multi-null topologies after a shot, and real-time reconstructors already
support selected boundary and primary-X quantities on several devices
\cite{anand2024,shi2026exl}.  For XPT, the enabling control requirement is
therefore not reconstruction in isolation, but \emph{closing the loop on the
secondary null}: keeping that null on the divertor leg of the primary
separatrix so that the intended flux expansion and magnetically decoupled
dissipation volume survive disturbances and shot-to-shot drift.  Absent such
feedback, secondary-null placement remains a feedforward commissioning task
even when $I_p$ and the global boundary track well---as illustrated by the
MAST-U residual field offset and multi-discharge tuning
\cite{anand2024}.

\begin{table}[htbp]
\centering
\caption{Experimental XPT magnetic control to date.  ``Real-time secondary
null in FB'' asks whether a real-time estimate of the secondary null enters
the feedback law (as opposed to feedforward coil programming alone).}
\label{tab:xpt-control}
\footnotesize
\setlength{\tabcolsep}{3.5pt}
\begin{tabular}{@{}lllc@{}}
\toprule
Device & Method & FB targets &
Real-time secondary null in FB \\
\midrule
TCV \cite{theiler2017,raj2022,lee2025,winkel2026} &
Divertor shape control &
$I_p$, boundary / shape &
No \\
MAST-U \cite{anand2024,lonigro2026} &
LEMUR + FF $B_{r,z}\!\to\!0$ &
$R_{\mathrm{OUT}}$, $R_{\mathrm{IN}}$, $Z_X$ &
No \\
EXL-50U (this exp.) &
FF+PID &
$I_p$, $R$, $Z$ &
No \\
\bottomrule
\end{tabular}\\[0.4em]
\end{table}

Regulating multi-null divertor geometry therefore requires one of two
interfaces.  The first closes feedback on explicit null states---null
positions and related flux constraints obtained from a real-time equilibrium
reconstruction or a dedicated null observer---together with $I_p$ and the last
closed flux surface (LCFS).  The second is reconstruction-free: a controller
maps magnetic probes, flux loops, and coil currents directly to actuator
commands, so that multi-null geometry is regulated without feeding a real-time
equilibrium into the loop.  Deep RL on TCV has demonstrated the latter for
snowflake and Jellyfish targets, including a snowflake transition that reduced
X-point separation from $34$ to $6.6$\,cm and the three-null Jellyfish
configuration, with X-point structure latent in the learned policy rather than
exposed as reconstructed feedback channels
\cite{degrave2022,tracey2024,gorno2024}.  Related reconstruction-free magnetic
control has also been shown on DIII-D and WEST
\cite{subbotin2026,kerboua2024west}.  Those results establish that multi-null
feedback need not be reconstruction-based; they do not, however, close an XPT
loop in which a \emph{secondary} null on the divertor leg is an explicit
real-time feedback variable.  On EXL-50U the reconstruction-free magnetics
interface for multi-null divertor control has not yet been validated, whereas
the planned plant stack routes equilibrium features from the real-time
reconstruction code PTEFIT to the controller
\cite{zheng2026ptefit,shi2026exl}.  We therefore first develop the explicit,
reconstruction-based route: closed-loop XPT control on reconstructed secondary-
null position and flux together with $I_p$ and the LCFS---an interface that,
to our knowledge, has not been demonstrated in XPT experiments to date.  We
do not claim superiority of either interface; the present choice matches the
near-term EXL-50U control architecture and makes the secondary-null channels
auditable for multi-objective learning.
Model-predictive control provides a non-learning alternative for shape
regulation \cite{mele2025mpc}, but a validated control-oriented model of the
coupled EXL-50U XPT nulls is not presently available.

XPT control is multi-objective on a single trajectory: plasma current, the
LCFS, and multiple null position/flux channels must remain jointly acceptable
under shared PF actuators, while the identity of the worst channel can change
over the discharge.  Fusion RL typically scalarizes these channels in the
reward before estimating one advantage
\cite{degrave2022,tracey2024}.  In robotics and multi-task learning, related
conflicts are instead attacked inside the learner---notably by
projected conflicting gradients (PCGrad)
\cite{yu2020pcgrad,munn2025gcr} and by
preference-conditioned policies with multi-head advantage decomposition
\cite{ambadkar2026d3po}.

Those methods mainly mitigate gradient interference or support preference
trade-offs across tasks.  Reconstructed-null XPT control is harder-coupled: a
weak null/flux channel can fail the topology even when the scalar return looks
acceptable, and tokamak magnetic RL still rarely keeps objective structure
through credit assignment.  The relevant design choice is therefore where
nonlinear scalarization occurs relative to objective-wise temporal credit.

The gap addressed here is therefore twofold.  At the control-system level,
existing XPT operation does not close feedback on reconstructed secondary-null
position and flux together with the plasma current and boundary.  At the
algorithmic level, conventional reward-level scalarization discards objective
identity before temporal credit assignment.  We develop AdvA along
the second axis: a shared critic backbone with one value head and one
generalised advantage estimation (GAE) \cite{schulman2016gae}
per channel, with worst-objective-aware
softmin aggregation (implementation name SmoothMax; $\alpha<0$) applied only
after advantage estimation, augmented by a controlled component residual on
the aggregated update.  On EXL-50U, PTEFIT already provides real-time
equilibrium reconstruction for boundary quantities at the centimetre level
\cite{zheng2026ptefit,shi2026exl}; whether its secondary-null position and
flux estimates are accurate enough for closed-loop XPT feedback remains to be
established experimentally.  The present study therefore validates the control
interface and AdvA in an experiment-calibrated free-boundary simulation, as a
precursor to PTEFIT-in-the-loop tests.

The contributions address these gaps as follows:
\begin{enumerate}
    \item We establish an AI-enabled multi-objective framework for closed-loop
    XPT magnetic configuration control in an experiment-calibrated EXL-50U
    free-boundary simulation environment.  The framework jointly regulates
    plasma current, LCFS geometry, X-point positions, and X-point flux
    constraints through shared poloidal-field actuators.
    \item We develop Advantage Aggregation (AdvA) for coupled multi-objective
    magnetic control.  By retaining objective-specific value heads and generalised
    advantage estimates before nonlinear scalarisation, together with a
    controlled residual correction, AdvA preserves objective-wise temporal
    credit assignment and supports coordinated policy optimisation.
    \item We evaluate the framework against conventional reward-level RL and
    the experiment-derived FF+PID baseline across nominal operation,
    measurement uncertainties, and unseen initial equilibria, together with
    multi-initial adaptation.  The results characterise its control capability,
    robustness, transferability, and remaining challenges for simulation-based
    XPT magnetic control.
\end{enumerate}
The present evidence is simulation-based; closed-loop XPT control on
EXL-50U with PTEFIT-supplied null states remains future experimental work.

\section{Methods}

\subsection{EXL-50U XPT control problem}
\label{sec:xpt-control-problem}

\begin{figure}[htbp]
    \centering
    \includegraphics[width=0.8\linewidth]
    {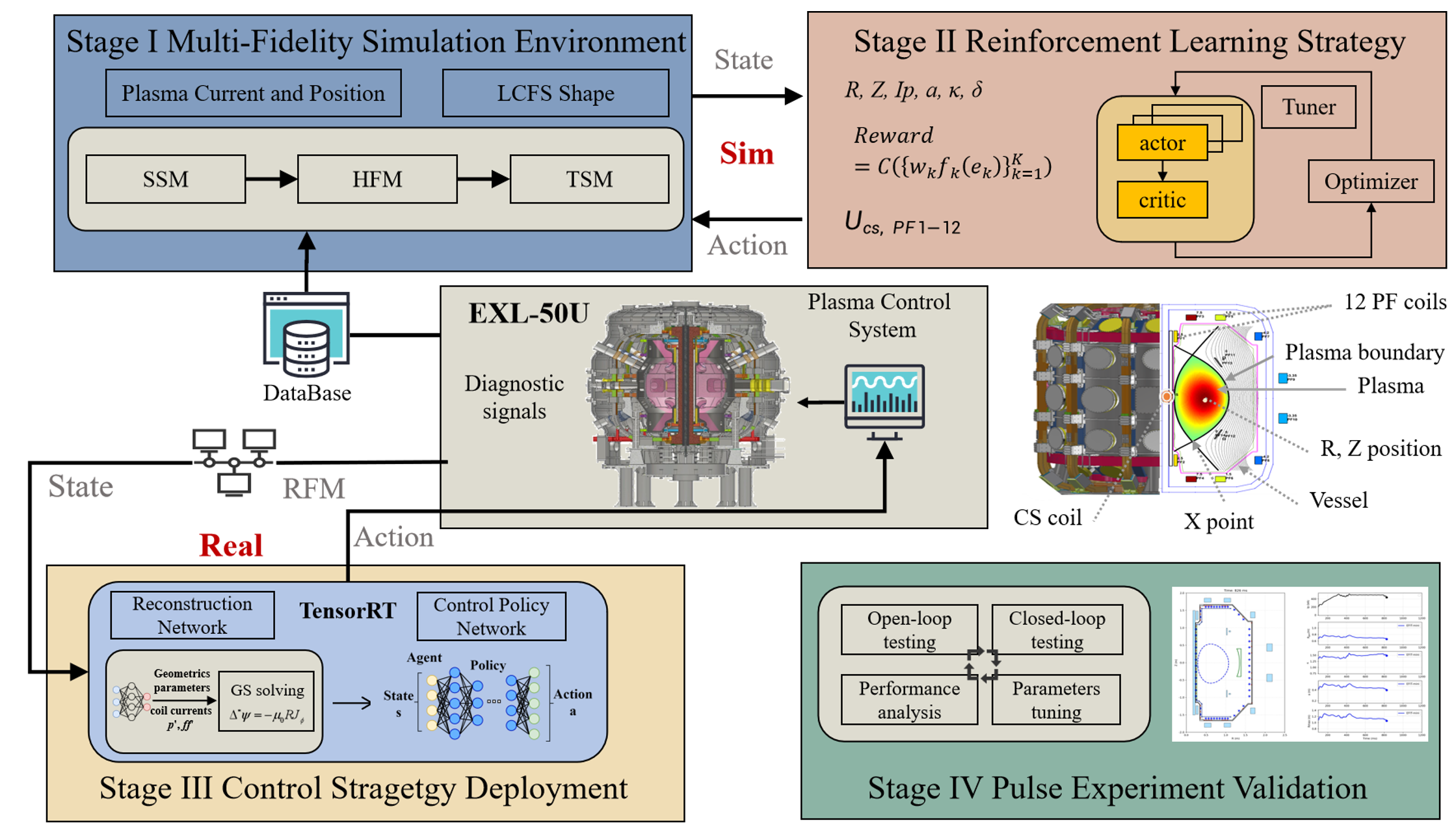}
    \caption{End-to-end architecture for reconstructed-null XPT control.
    Simulation closes the loop through FGE; deployment swaps FGE
    observations for real-time reconstruction while keeping the same
    state and coil-command interfaces (11 policy coils; VS under PID).}
    \label{fig:overall-framework}
\end{figure}

Figure~\ref{fig:overall-framework} summarises the EXL-50U control stack.
Twelve active coil circuits---the central solenoid (CS), ten Poloidal Field coils
(PF1--PF10), and the vertical-stability coil (VS)---regulate the plasma
current and magnetic configuration.  Learned policies command CS and
PS1--PS10 (11 voltages); VS remains under a separate PID loop due to highe frequency for every
controller reported here.  The target XPT has two primary nulls near the
confined boundary and two secondary nulls in the divertor legs.  We use
one-based labels $X_1$--$X_4$: $X_2$, $X_3$
are the upper/lower inner primary pair and $X_1$, $X_4$ the upper/lower
outer secondary pair.  The plant is strongly coupled---each coil
simultaneously affects $I_p$, the LCFS, and all four nulls---so improving
one X-point condition can displace another null or deform the boundary.

Training and evaluation close the loop through the calibrated fast
free-boundary Grad--Shafranov evolutive solver (FGE)
\cite{heiss2026fge} (Sec.~\ref{sec:fge-validation}): FGE supplies $I_p$,
coil currents, LCFS, and X-point features; the policy returns the 11 coil
voltages; the environment advances and scores the next step.  The same
observation and command interfaces are already wired for machine deployment
(Figure~\ref{fig:overall-framework}): diagnostic and coil signals pass
through reflective memory (RFM) to real-time equilibrium reconstruction,
the policy runs under TensorRT, and coil commands return to the power
supplies.  With that end-to-end path in place, a controller validated in
FGE is ready for on-machine closed-loop tests; the experiments below
remain simulation-based and do not yet report PTEFIT-in-the-loop XPT
feedback on EXL-50U.

\subsection{FGE simulation environment}

\subsubsection{Fast free-boundary Grad--Shafranov evolutive solver}
All controllers interact with FGE \cite{heiss2026fge}, a
control-oriented free-boundary equilibrium evolution solver in the Matlab
EQuilibrium suite (MEQ), previously used for RL magnetic control on TCV
\cite{degrave2022}.  Given PF-coil voltages, FGE advances the
axisymmetric equilibrium on resistive timescales by coupling three
blocks.  The poloidal flux satisfies the Grad--Shafranov equation
\begin{equation}
    \Delta^{*}\psi
    = -2\pi R\mu_0 j_\phi
    = -4\pi^{2}\left(\mu_0 R^{2} p'(\psi)+TT'(\psi)\right),
    \label{eq:fge_gs}
\end{equation}
with free profiles $p'(\psi)$ and $TT'(\psi)$ fixed by scalar
constraints (here $\beta_p$ and $q_0$).  Active and passive conductor
currents $I_e$ obey the circuit equation
\begin{equation}
    \begin{pmatrix} V_a \\ 0 \end{pmatrix}
    = M_{ee}\dot I_e + M_{ey}\dot I_y + R_e I_e,
    \label{eq:fge_circuit}
\end{equation}
where $V_a$ are the applied coil voltages, $I_y$ is the plasma current
distribution, and $M_{ee}$, $M_{ey}$, $R_e$ are the mutual-inductance and
resistance matrices.  The bulk plasma current is closed by a rigid Ohmic
model (OhmTor-rigid in \cite{heiss2026fge}),
\begin{equation}
    0 = L_p\dot I_p + M_{pe}\dot I_e + R_p\left(I_p - I_{ni}\right),
    \label{eq:fge_cde}
\end{equation}
with plasma self-inductance $L_p$, plasma--conductor mutual $M_{pe}$,
bulk resistance $R_p$, and non-inductive current $I_{ni}$.  Each step
returns $I_p$, coil currents, the LCFS, X-point positions, and fluxes
for observations and rewards.  On EXL-50U we use a $66\times65$ $(R,Z)$
grid, 544 passive filaments plus 12 active circuits, and a fixed
$1\,\mathrm{ms}$ step; $\beta_p$ and $R_p$ are held at
discharge-calibrated values (Sec.~\ref{sec:fge-validation}).  The model
is restricted to the magnetic degrees of freedom needed for XPT topology
regulation and is fast enough for parallel policy training.

\subsubsection{Calibration and validation on \#13906}
\label{sec:fge-validation}
The environment is calibrated to EXL-50U discharge \#13906, which sustained
an XPT during the $I_p$ flat top near $400$--$700\,\mathrm{ms}$.
Reference equilibria come from the LIUQE tokamak equilibrium reconstruction
code \cite{moret2015liuqe}; we use the $500$--$700\,\mathrm{ms}$ window.
Shot-specific inputs are the reconstructed $\beta_p(t)$ and $q_0(t)$ as
profile constraints in Eq.~\eqref{eq:fge_gs}, and a constant bulk plasma
resistance $R_p$ chosen so that simulated $I_p$ tracks the experiment over
that window.  Coil and vessel geometry, inductances, resistances, and
power-supply parameters follow engineering calibrations (including vacuum
shots) and are held fixed.

\begin{figure}[htbp]
    \centering
    \includegraphics[width=0.95\textwidth]{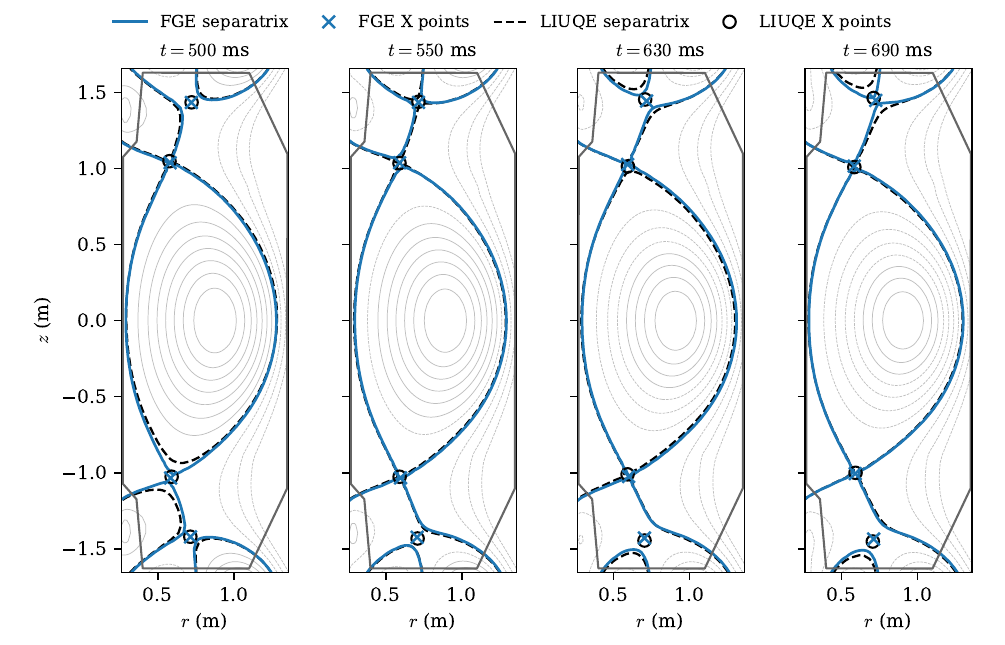}
    \caption{Free-boundary equilibria from the FGE validation run of
    EXL-50U discharge \#13906 at $t=0.50$, $0.55$, $0.63$, and
    $0.69\,\mathrm{s}$ (left to right).  Each panel
    overlays the simulated separatrix (solid blue) and X points (blue
    crosses) on the experimental LIUQE-reconstructed separatrix (dashed
    black) and X points (open circles); thin gray curves show simulated
    flux surfaces and the dark gray contour marks the limiter.}
    \label{fig:fge_equilibria}
\end{figure}

\begin{figure}[htbp]
    \centering
    \includegraphics[width=0.95\textwidth]{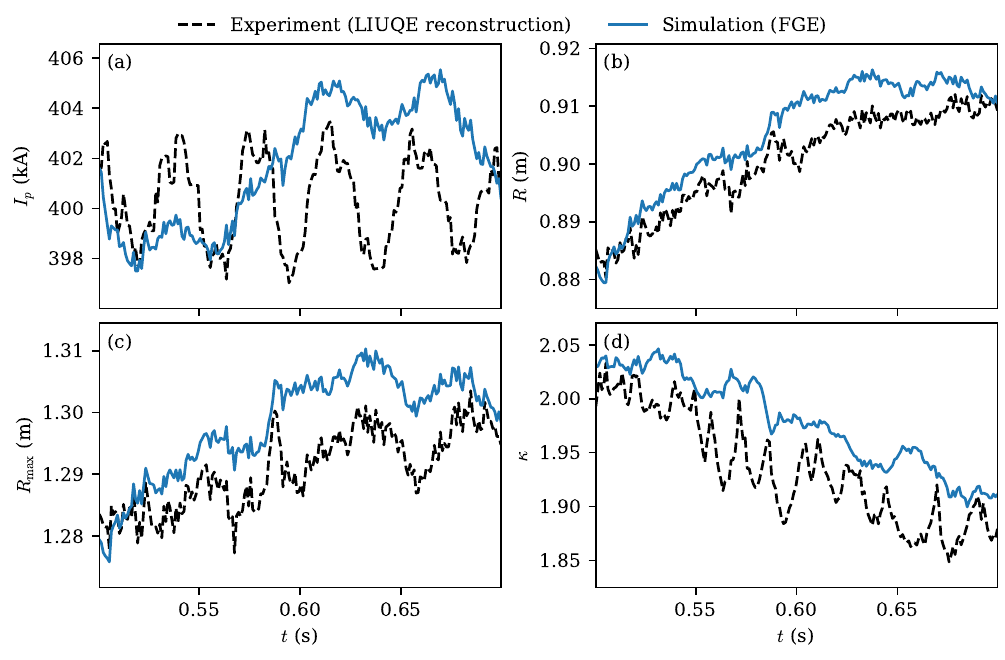}
    \caption{Time traces from the FGE validation run of EXL-50U
    discharge \#13906 (solid blue) compared with the experimental LIUQE
    reconstruction (dashed black): (a) plasma current $I_p$;
    (b) plasma radial position $R$;
    (c) outer radial extent $R_{\max}$ of the LCFS; (d) LCFS elongation
    $\kappa$.}
    \label{fig:fge_traces}
\end{figure}

For validation, FGE starts from the LIUQE equilibrium at
$t=0.5\,\mathrm{s}$ and advances to $t=0.7\,\mathrm{s}$ under the
recorded coil voltages.  Figure~\ref{fig:fge_equilibria} overlays
boundaries and nulls;
Figure~\ref{fig:fge_traces} compares $I_p$, $R$, $R_{\max}$, and $\kappa$;
Table~\ref{tab:fge_validation} summarises bias, RMSE, and maxima.
The XPT topology (all four nulls) is preserved; $I_p$ RMSE is
$0.8\%$ ($1.6\%$ at worst), and LCFS/X-point offsets stay at the
centimetre scale of the grid cell ($\approx1.7\,\mathrm{cm}$), which
matches the tolerances of the configuration-control task.  Residual
mismatches concentrate in the divertor legs and in small biases of
$\kappa$ and $R_{\max}$, consistent with the reduced profile model.

\begin{table}[htbp]
    \centering
    \caption{Deviations between the FGE validation run and the LIUQE
    reconstruction of discharge \#13906 over $0.5$--$0.7\,\mathrm{s}$:
    time-averaged signed deviation (bias, simulation minus
    reconstruction), root-mean-square error (RMSE), and maximum
    absolute deviation.  The LCFS row is based on the symmetrized mean
    distance between the simulated and reconstructed boundary contours,
    and the X-point rows on the Euclidean position offsets averaged
    over the upper and lower X point of each type; these distances are
    nonnegative by construction, so no bias is reported.}
    \label{tab:fge_validation}
    \begin{tabular}{lccc}
        \toprule
        Quantity & Bias & RMSE & Maximum \\
        \midrule
        $I_p$ (kA) & $+1.4$ & 3.1 & 6.5 \\
        $R$ (mm) & $+4.7$ & 5.3 & 11.4 \\
        $R_{\max}$ (mm) & $+7.6$ & 8.7 & 16.7 \\
        $\kappa$ & $+0.045$ & 0.051 & 0.101 \\
        LCFS distance (mm) & --- & 13.8 & 30.5 \\
        Primary X points $X_2$, $X_3$ (mm) & --- & 18.3 & 38.0 \\
        Secondary X points $X_1$, $X_4$ (mm) & --- & 13.5 & 21.6 \\
        \bottomrule
    \end{tabular}
\end{table}

\FloatBarrier

\subsection{Control methods}

\subsubsection{Baseline: FF+PID}
\label{sec:baseline-ff-pid}

\begin{figure}[htbp]
    \centering
    \includegraphics[width=0.8\linewidth]
    {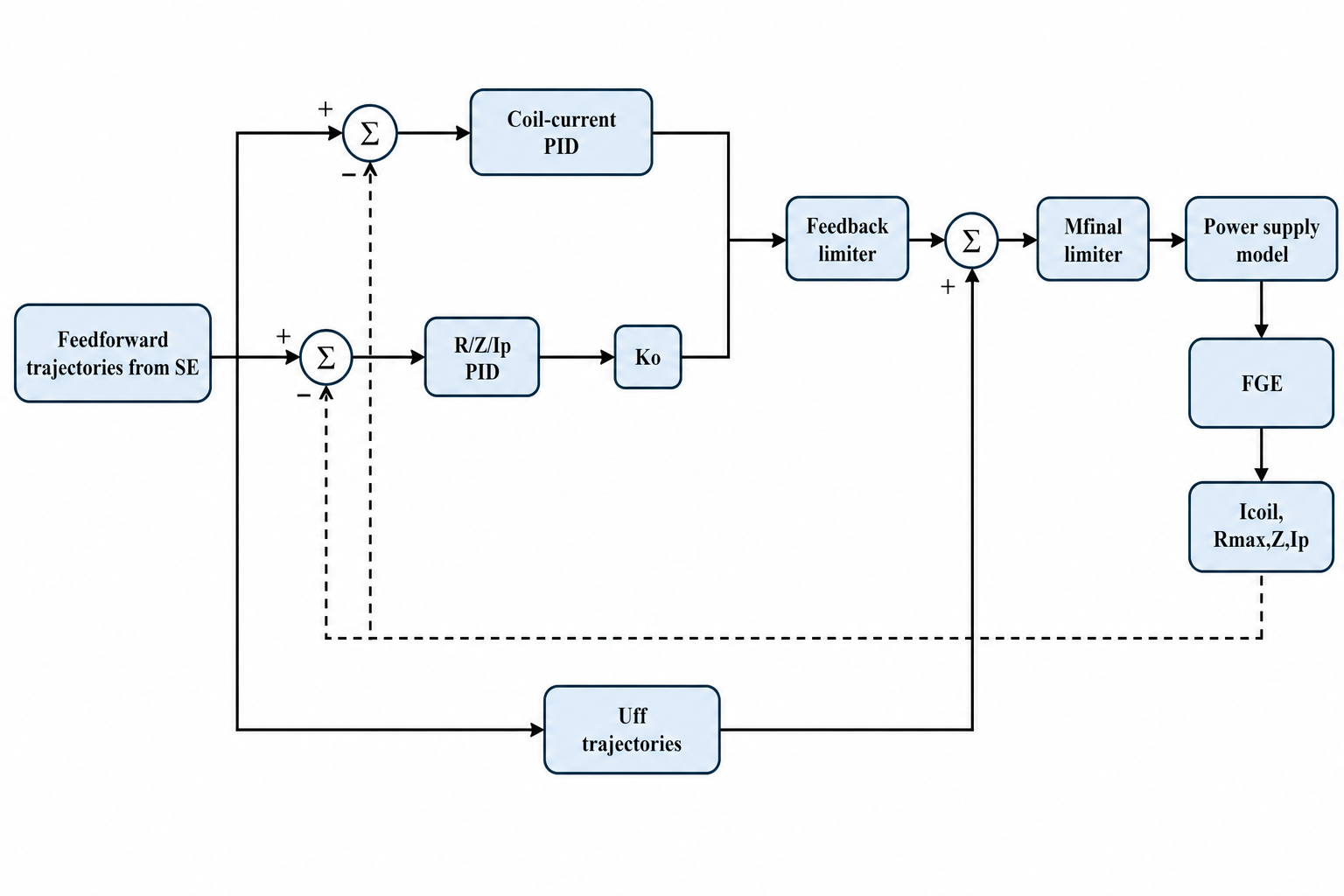}
    \caption{Baseline FF+PID control structure.}
    \label{fig:baseline-ff-pid}
\end{figure}

The classical baseline is feedforward plus PID (Figure~\ref{fig:baseline-ff-pid}).
Offline, the MATLAB Shape Editor (SE) designs PF-coil current and $I_p$
references and the associated feedforward voltages
\cite{song2019hl2m}.  Online, PID loops on PF-coil currents, $I_p$, and
plasma position $(R,Z)$ produce a voltage correction that is added to the
feedforward,
\begin{equation}
    \boldsymbol{V}_{\mathrm{cmd}}
    =
    \boldsymbol{V}_{\mathrm{ff}}
    +
    \Delta\boldsymbol{V}_{\mathrm{PID}}.
\end{equation}
All FF+PID scores in this paper reuse the \#13906 plant feedforward and
PID gains in a closed-loop FGE rollout, without per-initial redesign of
the feedforward table.

\subsubsection{RL control}
\label{sec:rl-control}

We cast XPT magnetic control as a partially observed decision process:
the full FGE equilibrium is the latent state, but the feedforward policy
$\pi(\mathbf{a}_t\mid\mathbf{o}_t)$ acts only on a reconstructed
observation $\mathbf{o}_t$ and is trained to maximise the discounted return
$\mathbb{E}_\pi[\sum_t\gamma^t r_t]$.  The interface used throughout is as
follows.

\paragraph{Observation:} The 45-dimensional observation $\mathbf{o}_t$
    concatenates normalised plasma-current error, 12 coil currents
    (CS, PS1--PS10, and VS), 4 X-point features (validity, $R$--$Z$ position
    errors, and flux relative to the LCFS), and radial and vertical LCFS
    errors at eight boundary points equally spaced in poloidal angle about
    the magnetic axis.  The same observation is used for every learned
    controller in this paper.

\paragraph{Action:} The learned policy commands an 11-dimensional coil-voltage
    vector (CS and PS1--PS10).  The vertical-stability coil (VS) is regulated
    by a separate PID loop and is excluded from the policy action for every
    controller in this paper.  Policy outputs are passed through $\tanh$
    squashing and rescaled to the physical voltage limits.  Actuation metrics
    $|\Delta V|$ are computed on these 11 policy-commanded coils.

\paragraph{Reward:} Tracking errors for the plasma current, the four
    X-point positions, the four X-point flux-difference conditions, and the
    LCFS boundary are extracted from the FGE output and mapped to
    satisfaction signals $r_{t,i}\in[0,1]$ by sigmoid or softplus shaping
    with ``good''/``bad'' thresholds.  How these $K$ signals are aggregated
    relative to temporal credit assignment is the algorithmic axis of the
    next subsections.

\subsubsection{Actor--critic instantiation with PPO}
\label{sec:ppo}

Advantage Aggregation (AdvA) needs a per-objective advantage signal, so it
belongs to the actor--critic family: methods that maintain a critic (or
critics) and form advantages for a policy update.  Pure value-based
algorithms without an explicit advantage pathway, such as the deep
Q-network (DQN) \cite{mnih2015dqn}, are outside this scope.  We instantiate
both the reward-level baseline and AdvA with PPO
\cite{schulman2017ppo}; the same aggregation design can be attached to
other actor--critic algorithms.  In results we write \textbf{AdvA-PPO} for
the AdvA instance on PPO.

PPO stabilises policy-gradient updates through a clipped surrogate.
With probability ratio
$\chi_t(\theta)=\pi_\theta(\mathbf a_t\mid\mathbf o_t)/
\pi_{\theta_{\mathrm{old}}}(\mathbf a_t\mid\mathbf o_t)$
and advantage $\widehat A_t$,
\begin{equation}
    \mathcal L^{\mathrm{CLIP}}(\theta)
    = -\mathbb{E}_t\!\left[
        \min\!\left(
            \chi_t(\theta)\,\widehat A_t,\;
            \operatorname{clip}\!\bigl(
                \chi_t(\theta), 1-\epsilon, 1+\epsilon
            \bigr)\,\widehat A_t
        \right)
    \right],
    \label{eq:ppo_clip}
\end{equation}
the clip removes the incentive for large probability-ratio steps when they
would inflate the objective.  The full training loss adds a value-function
error and an entropy bonus,
\begin{equation}
    \mathcal L^{\mathrm{PPO}}(\theta)
    = \mathcal L^{\mathrm{CLIP}}(\theta)
    + c_1\,\mathcal L^{\mathrm{VF}}(\theta)
    - c_2\,\mathcal H[\pi_\theta],
    \label{eq:ppo_full}
\end{equation}
where $\mathcal L^{\mathrm{VF}}$ fits the critic to empirical returns and
$\mathcal H$ encourages exploration.  Both the reward-level baseline and
every AdvA variant share the $45$-dimensional observation and
$11$-dimensional action of Sec.~\ref{sec:rl-control}.

\subsubsection{Reward-level SmoothMax baseline}
\label{sec:reward-ppo}

The conventional multi-objective baseline scalarises the channel rewards
\emph{before} GAE \cite{schulman2016gae}: that is why we call
it reward-level.  The same worst-objective-aware exponential weighting used
later by AdvA defines
\begin{align}
    \omega_{t,i}
    &=
    \frac{w_i^0\exp(\alpha r_{t,i})}
    {\sum_j w_j^0\exp(\alpha r_{t,j})},\\
    u(\mathbf r_t)=\bar r_t
    &=
    \sum_i \omega_{t,i}r_{t,i},
    \qquad \alpha<0 ,
    \label{eq:smoothmax_anchor_reward}
\end{align}
so a poorly satisfied channel receives more weight.  We keep the
implementation name ``SmoothMax''; with $\alpha<0$ the softmin emphasises
the worst objective.  A single value function and GAE fitted to $\bar r_t$
then supply the PPO advantage in Eq.~\eqref{eq:ppo_clip}.  Results and
ablations refer to this baseline as \textbf{Reward-PPO}
(Table~\ref{tab:controller-inventory}).

\subsubsection{Advantage Aggregation (AdvA)}
\label{sec:adva}

\begin{figure}[htbp]
    \centering
    \includegraphics[width=\linewidth]
    {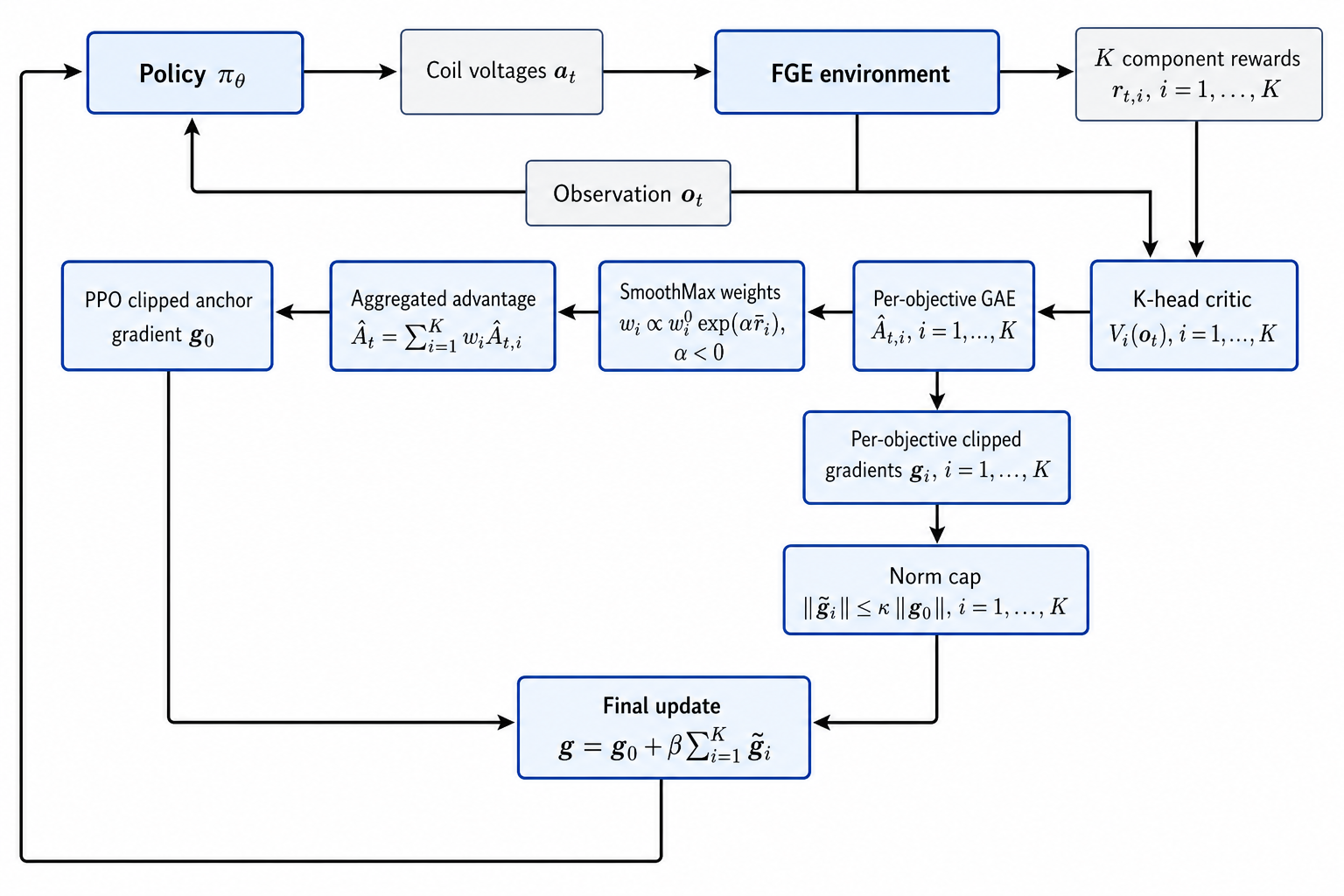}
    \caption{Policy update and environment interaction for AdvA-PPO.
    The FGE environment provides the reconstructed magnetic observation and
    component rewards.  Per-objective value heads and GAE preserve temporal
    credit assignment before SmoothMax aggregation; the PPO anchor is then
    corrected by the norm-capped residual of
    Eqs.~\eqref{eq:adva_norm_cap}--\eqref{eq:adva_total_gradient}.
    Reliability gating and gradient projection are ablation options and are
    off in AdvA-PPO.}
    \label{fig:adva-pipeline}
\end{figure}

AdvA moves scalarisation to the \emph{advantage} layer: one value head and
one GAE per physical channel on a shared critic backbone, with nonlinear
mixing only afterwards.  Relative
to reward-level SmoothMax, a weak channel no longer collapses the temporal
credit of every other channel before advantage estimation.
Figure~\ref{fig:adva-pipeline} shows the closed loop and update path.

Let $r_{t,i}\in[0,1]$ be the channel reward and $w_i^{0}$ a fixed prior
weight (inner X-point positions and the LCFS receive larger $w_i^{0}$).
A shared critic backbone feeds heads $V_i(\mathbf o_t)$ with
\begin{align}
    \delta_{t,i}
    &= r_{t,i}+\gamma V_i(\mathbf o_{t+1})-V_i(\mathbf o_t),\\
    \widehat A_{t,i}
    &= \sum_{\ell=0}^{T-t-1}(\gamma\lambda)^\ell
    \delta_{t+\ell,i},
    \qquad i=1,\ldots,K .
    \label{eq:objective_wise_gae}
\end{align}
Each head is trained against
$\widehat G_{t,i}=\widehat A_{t,i}+V_i(\mathbf o_t)$.

\paragraph{SmoothMax advantage aggregation.}
Batch-mean channel rewards $\bar r_i$ define worst-case-aware weights
\begin{equation}
    w_i
    =
    \frac{w_i^{0}\exp(\alpha\bar r_i)}
    {\sum_{j}w_j^{0}\exp(\alpha\bar r_j)},
    \qquad \alpha<0,
    \label{eq:adv_smoothmax_weights}
\end{equation}
and the scalar advantage
\begin{equation}
    \widehat A_t
    =
    \sum_{i=1}^{K} w_i\,\widehat A_{t,i}
    \label{eq:adv_smoothmax_aggregate}
\end{equation}
replaces the usual PPO advantage column.  Setting $\alpha=0$ recovers a
fixed-weight mean (Adv-mean in the ablation).  The actor uses a single
clipped surrogate on $\widehat A_t$.

\paragraph{Reliability gate.}
A head with low explained variance, or a channel already saturated near
reward~$1$, contributes little useful gradient.  Exponential moving
averages of explained variance $\overline{\mathrm{EV}}_i$ and saturation
fraction $\bar s_i$ define
\begin{equation}
    q_i
    =
    \operatorname{clip}
    \left(\overline{\mathrm{EV}}_i,0,1\right)
    (1-\bar s_i)
    \in[0,1],
    \label{eq:adva_reliability_gate}
\end{equation}
and when enabled multiply the aggregation weights,
$w_i\leftarrow w_i q_i$, followed by renormalisation.  The gate is an
ablation option and is off in AdvA-PPO.

\paragraph{Norm-capped residual.}
PPO clipping acts once on the aggregated advantage, so a minority channel
whose $\widehat A_{t,i}$ disagrees with $\widehat A_t$ can be silenced.
Component surrogates
\begin{equation}
    \mathcal{L}_{i}^{\mathrm{clip}}(\theta)
    =
    -\mathbb{E}_t
    \left[
    \min\left(
    \chi_t(\theta)\widehat A_{t,i},
    \operatorname{clip}
    \bigl(\chi_t(\theta),1-\epsilon,1+\epsilon\bigr)
    \widehat A_{t,i}
    \right)
    \right]
    \label{eq:objective_ppo_surrogate}
\end{equation}
supply gradients
$\mathbf g_0=\nabla_\theta\mathcal{L}^{\mathrm{CLIP}}(\widehat A)$
and
$\mathbf g_i=\nabla_\theta\mathcal{L}_{i}^{\mathrm{clip}}$.
Each component is capped to the anchor scale,
\begin{equation}
    \widetilde{\mathbf g}_i
    =
    \mathbf g_i
    \min\left(
    1,\frac{\kappa\lVert\mathbf g_0\rVert_2}
    {\lVert\mathbf g_i\rVert_2+\varepsilon_g}
    \right),
    \label{eq:adva_norm_cap}
\end{equation}
and the update is
\begin{equation}
    \mathbf g
    =
    \mathbf g_0+\beta\sum_{i=1}^{K}
    \widetilde{\mathbf g}_i ,
    \label{eq:adva_total_gradient}
\end{equation}
with defaults $\kappa=1$ and $\beta=0.2$.  Setting $\beta=0$ recovers pure
advantage-level SmoothMax.

\paragraph{Gradient projection.}
As an optional conflict-resolution step in the spirit of PCGrad
\cite{yu2020pcgrad,munn2025gcr}, the component of
$\widetilde{\mathbf g}_i$ opposing $\mathbf g_0$ can be removed before the
sum in Eq.~\eqref{eq:adva_total_gradient}.  Projection appears only in
proxy ablation rows of Sec.~\ref{sec:mechanism-results} and is disabled in
AdvA-PPO ($\mathbf p_i=\widetilde{\mathbf g}_i$).

\paragraph{Proposed controller: AdvA-PPO.}
\textbf{AdvA-PPO} is the AdvA instance used in all main experiments:
advantage-level SmoothMax ($\alpha=-3$) plus the capped residual
($\beta=0.2$, $\kappa=1$), with the reliability gate and projection both
off.  Sec.~\ref{sec:mechanism-results} motivates retaining the residual and
omitting gate and projection.

\subsubsection{Ablation sequence and controller inventory}
\label{sec:controller-inventory}
Table~\ref{tab:controller-inventory} lists the controllers compared in the
Results ablation (Sec.~\ref{sec:mechanism-results}).  Results name the
reward-level SmoothMax PPO baseline of Sec.~\ref{sec:reward-ppo} as
\textbf{Reward-PPO}.  The ablation follows the design decisions in causal
order:
\begin{enumerate}
    \item Start from Reward-PPO (reward-level SmoothMax, scalar GAE).
    \item Move to Adv-mean ($\alpha{=}0$, multi-head).
    \item Replace the mean by Adv-SmoothMax ($\alpha{=}-3$, multi-head).
    \item On that SmoothMax anchor, compare $+\;$gate ($\beta{=}0$)
    with $+\;$resid ($\beta{=}0.2$) $=$ AdvA-PPO.
\end{enumerate}
Pairwise gate/projection combinations are reported only as temporary proxies
from available checkpoints, not as a second method family.
Early multi-head runs still carry two radial-extrema rewards
($R_{\min}$, $R_{\max}$), hence $K=12$; those channels are excluded from
every reported score.  Clean contrasts in Sec.~\ref{sec:mechanism-results} are
therefore Adv-mean versus Adv-SmoothMax ($K=12$), gate versus residual on
the SmoothMax anchor ($K=10$), and the family-level Reward-PPO versus
AdvA-PPO comparison; proxy rows are not matched one-variable AdvA-native
cells.  Within a channel set, matched variants share the $45/11$
observation--action interface, capacity, optimizer, and checkpoint rule;
FF+PID is evaluated, not retrained.

\begin{table}[htbp]
\centering
\caption{Controllers in the component ablation, in the same order as the
Results ablation (Sec.~\ref{sec:mechanism-results}).  ``Locus'' is where
nonlinear aggregation occurs
relative to objective-wise temporal credit.  For AdvA rows, Gate/$\beta$/Proj.\
act on advantage weights or the residual of
Eqs.~\eqref{eq:adva_norm_cap}--\eqref{eq:adva_total_gradient}.
All rows share the same $d_a{=}11$ / $d_o{=}45$ interface
(Sec.~\ref{sec:rl-control}; VS under PID).
\textbf{Reward-PPO} is the reward-level SmoothMax baseline of
Sec.~\ref{sec:reward-ppo}.}
\label{tab:controller-inventory}
\scriptsize
\setlength{\tabcolsep}{2pt}
\begin{tabular}{@{}llllcccc@{}}
\toprule
Name & Locus & $\alpha$ & Gate & $\beta$ & Proj. & $K$ & $d_a/d_o$ \\
\midrule
Reward-PPO & reward & $-3$ & off & 0 & off & 10 & $11/45$ \\
Adv-mean & advantage & $0$ & off & 0 & off & 10 & $11/45$ \\
Adv-SmoothMax & advantage & $-3$ & off & 0 & off & 10 & $11/45$ \\
$+$ gate & advantage & $-3$ & on & 0 & off & 10 & $11/45$ \\
$+$ resid (AdvA-PPO) & advantage & $-3$ & off & $0.2$ & off & 10 & $11/45$ \\
$+$ proj & advantage & $-3$ & off & $0$ & on & 10 & $11/45$ \\
$+$ gate$+$resid & advantage & $-3$ & on & $0.2$ & off & 10 & $11/45$ \\
$+$ gate$+$proj & advantage & $-3$ & on & $0$ & on & 10 & $11/45$ \\
\bottomrule
\end{tabular}
\end{table}

\paragraph{Training and inference implementation.}
The actor and shared value backbone are two fully connected layers of 256
units with $\tanh$ activations; AdvA controllers attach one scalar value head
per reward channel.  Policies use a squashed Gaussian action distribution and
are trained in PyTorch with Adam at learning rate $3\times10^{-4}$.
The PPO clip is $\epsilon=0.2$, $\gamma=0.98$, $\lambda=0.95$, the entropy
coefficient is $5\times10^{-3}$, and global gradient norm is clipped at 0.3.
Twenty-three parallel FGE workers collect one 300-step episode each per
iteration, giving a batch of 6900 transitions; each batch is optimised for ten
epochs with minibatches of 256.  The reported AdvA-PPO checkpoint is the final
checkpoint after 300 iterations ($2.07\times10^6$ environment steps).
Evaluation uses the deterministic mean action.

\FloatBarrier

\subsection{Evaluation setup}
\label{sec:eval-setup}
All reported scores come from one harness, one FGE image pool, and one
primary evaluation seed ($20260715$), with a $500$-step ($500\,\mathrm{ms}$)
closed-loop test horizon and deterministic actions at test time.
Training episodes are $300$ steps ($300\,\mathrm{ms}$); finishing the longer
test window is a temporal-extrapolation check. 

\subsubsection{Test scenarios}
\label{sec:eval-scenarios}
Four initial equilibria (I1--I4; Table~\ref{tab:eval-initials}) share the
training inductance basis and request the same frozen \#13906 XPT target.
Each control window starts near $t_0\approx 500\,\mathrm{ms}$ of the parent
discharge and runs for $500\,\mathrm{ms}$.  I1 is the calibrated XPT
training initial; I2--I4 are withheld cross-initialization cells (same
geometric target and reward operators; only the FGE initial and $I_p$
reference change).

Robustness cells R1--R3 reuse I1.  Disturbances enter the
\emph{controller observation} only; the simulator state used for scoring
stays clean.  R1 redraws an observation delay uniformly in
$1$--$3\,\mathrm{ms}$ at every control step (non-monotonic jitter, harder
than a fixed mean delay).  R2 adds full diagnostic/coil noise:
$I_p$ $\sigma{=}6\,\mathrm{kA}$, per-coil $I_{\mathrm{PF}}$ at $1\%$, and
absolute $5\,\mathrm{mm}$ on raw X-point and LCFS coordinates.  R3 applies
R1 and R2 together.

\begin{table}[htbp]
  \centering
  \small
  \caption{Evaluation initials I1--I4.  All windows start at
  $t_0\approx 500\,\mathrm{ms}$ (CSV start $0.502\,\mathrm{s}$) and use the
  frozen \#13906 XPT geometric target; $I_p^\star$ is the current reference
  for that cell.}
  \label{tab:eval-initials}
  \begin{tabular}{@{}clllc@{}}
    \toprule
    ID & Shot & $t_0$ & Topology & $I_p^\star$ \\
    \midrule
    I1 & \#13906 & $\approx 500\,\mathrm{ms}$ & XPT (training / test) & $\approx 400\,\mathrm{kA}$ \\
    I2 & \#13705 & $\approx 500\,\mathrm{ms}$ & divertor (test) & $\approx 500\,\mathrm{kA}$ \\
    I3 & \#13844 & $\approx 500\,\mathrm{ms}$ & divertor (test) & $\approx 500\,\mathrm{kA}$ \\
    I4 & \#15892 & $\approx 500\,\mathrm{ms}$ & limiter (test) & $\approx 500\,\mathrm{kA}$ \\
    \bottomrule
  \end{tabular}
\end{table}

\subsubsection{Metrics and statistical protocol}
\label{sec:eval-metrics}
Scores are recomputed from archived equilibria by one pipeline.  Raw FGE
X-point candidates are mapped to the four slots $X_1$--$X_4$ of
Sec.~\ref{sec:xpt-control-problem} before any error is formed.  Position
error is the Euclidean distance to the frozen target in that slot; flux
error is the poloidal-flux difference $|\psi_X-\psi_B|$ in that slot,
reported in Wb.  Mean $d_X$ or mean flux is the
average of the four per-slot RMSE\,/\,max values (primary/secondary columns
average only the corresponding pair).  The LCFS channel is the boundary
deviation used by the reward operator,
\begin{equation}
  e_{\mathrm{LCFS}}
  =
  \tfrac12
  \Bigl(
    \overline{\lvert (r_B-r_B^{\mathrm{ref}})/(\lvert r_B^{\mathrm{ref}}\rvert+\epsilon) \rvert}
    +
    \overline{\lvert (z_B-z_B^{\mathrm{ref}})/(\lvert z_B^{\mathrm{ref}}\rvert+\epsilon) \rvert}
  \Bigr),
  \qquad \epsilon=10^{-6}\,\mathrm{m},
\end{equation}
on eight boundary points equally spaced in poloidal angle about the
magnetic axis.  Tables report RMSE and max
absolute error over the survived horizon, survival length (full
$=500\,\mathrm{ms}$), and $|\Delta V|$ on the 11 policy coils.  On short
survivals, errors are secondary to survival; stochastic noise/jitter cells
are multi-seed averaged or labelled as a single realisation.

Because physical rows have incompatible units, we also report time means
of the shared ten-channel scores
($I_p$, four positions, four fluxes, LCFS): the reward-level SmoothMax
$\bar u$ of Eq.~\eqref{eq:smoothmax_anchor_reward} ($\alpha=-3$) and the
hard worst-channel mean
$\overline{r}=\overline{\min_i r_{t,i}}$.
Both are defined for FF+PID as well; evaluation $\bar u$ is not the
advantage-level SmoothMax inside AdvA.  Mechanism narrative emphasises
physical RMSE\,/\,max; Sec.~\ref{sec:mechanism-results} also lists $\bar u$ and
$\overline{r}$.  In comparison tables, the best entry in each column
(within a block) is \textbf{bold} and the second-best is underlined;
lower is better for errors and $|\Delta V|$, higher for survival,
$\bar u$, and $\overline{r}$.
\section{Results}

\subsection{Main controller comparison}
\label{sec:main-results}

Under the protocol of Sec.~\ref{sec:eval-setup}, the nominal cell uses the
calibrated shot-\#13906 initial (training initial; I1 of
Table~\ref{tab:eval-initials}) over a $500\,\mathrm{ms}$ rollout at
$1\,\mathrm{ms}$ per control step.  Three controllers are compared:
FF+PID, Reward-PPO, and AdvA-PPO (ours).  The head-to-head isolates the
aggregation locus (reward-level versus advantage-level); the component
ablation of Sec.~\ref{sec:mechanism-results} isolates the algorithmic
factors within AdvA.

\begin{figure}[htbp]
  \centering
  \includegraphics[width=\textwidth]{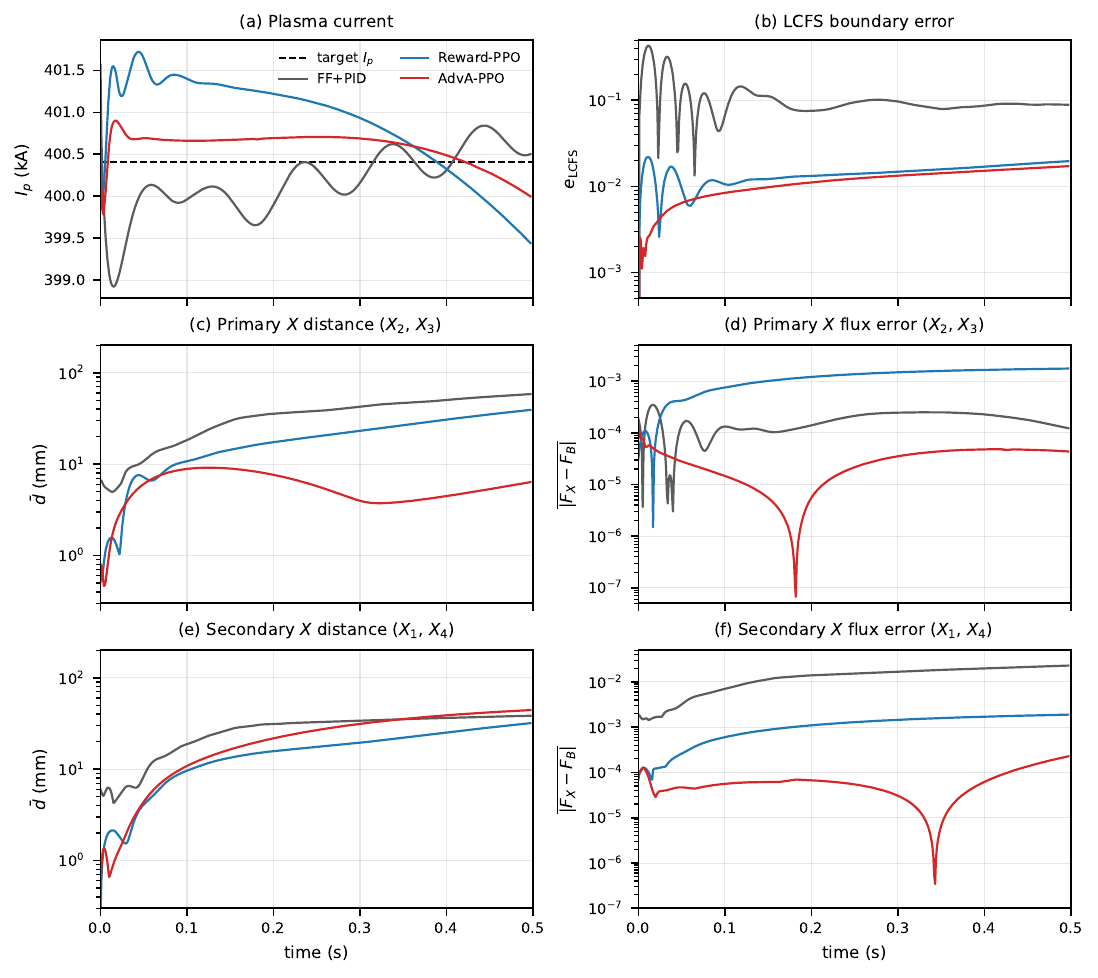}
  \caption{Closed-loop evolution on \#13906 ($500\,\mathrm{ms}$).
  Grey: FF+PID; blue: Reward-PPO; red: AdvA-PPO.
  (a)~$I_p$; (b)~LCFS error;
  (c)--(d)~primary $X$ distance and flux ($X_2$, $X_3$);
  (e)--(f)~secondary $X$ distance and flux ($X_1$, $X_4$).}
  \label{fig:main-comparison}
\end{figure}

Figure~\ref{fig:main-comparison} and Table~\ref{tab:main-comparison} show
that AdvA-PPO is the strongest multi-objective controller on this cell:
it leads or matches on $I_p$, primary $X$ distance, mean $X$ flux, and
LCFS, rather than trading one channel for another.  All three controllers
complete the horizon, so the comparison is tracking quality, not survival.
The evolution traces make the channel-wise behaviour visible in time;
panel~(d) already shows the primary-flux gap that the table later
quantifies as about a $20\times$ lower mean $X$-flux RMSE for AdvA-PPO
relative to Reward-PPO ($0.66$ versus $13.1\times10^{-4}$).

The decisive contrast is not a single physical row but the shared scoring
axis of Sec.~\ref{sec:eval-metrics}.  Reward-PPO reaches
$\bar u=0.56$ with worst-channel mean $\overline{r}=0.23$, limited by the
inner primary flux $X_2$; AdvA-PPO reaches $\bar u=0.93$ with
$\overline{r}=0.81$---a $3.5\times$ lift of the hard worst channel, together
with a $2.8\times$ lower $I_p$ RMSE ($260$ versus $735\,\mathrm{A}$).
Reward-PPO is not uniformly weaker: it records the lowest secondary $X$
distance ($19.0$ versus $28.2\,\mathrm{mm}$), but that gain coincides with
the poorest $I_p$ among the three and a much weaker primary-flux channel.

The secondary-null rows deserve a separate reading, because position and flux
are not interchangeable descriptors of an XPT.  The defining topological
condition is that the secondary null sits \emph{on} the divertor leg of the
primary separatrix, i.e.\ $\psi_X\!\to\!\psi_B$; where along that leg it sits
sets the location of the flux expansion but does not decide whether the
configuration is an XPT at all.  On this cell AdvA-PPO holds the secondary
flux condition an order of magnitude tighter than Reward-PPO
($1.07$ versus $17.1\times10^{-4}\,\mathrm{Wb}$ on $X_1$ and $0.61$ versus
$8.8\times10^{-4}\,\mathrm{Wb}$ on $X_4$) while allowing the nulls to sit
about $10\,\mathrm{mm}$ further along the leg.  In other words the two
policies fail differently: AdvA-PPO keeps the nulls locked to the separatrix
and slides them along it, whereas Reward-PPO places them closer to the
programmed coordinates but lets them detach from the leg.  For the XPT the
first failure mode is the benign one, and the residual $\sim\!10\,\mathrm{mm}$
offset is comparable to the secondary-null reconstruction accuracy of the
calibration itself (Table~\ref{tab:fge_validation}, $13.5\,\mathrm{mm}$
RMSE), so it should not be read as a resolved ranking.  The flux criterion is
also intrinsically weak against displacement: $\nabla\psi=0$ at a null, so
$|\psi_X-\psi_B|$ grows only quadratically with the null displacement and a
small flux error certifies topology rather than position.  The two channels
are therefore complementary, and we report both rather than a single
secondary-null figure of merit.
FF+PID keeps competitive $I_p$ ($454\,\mathrm{A}$ RMSE) because its loops
close on coil currents, $I_p$, and the geometric centroid ($R$, $Z$); it
does not feed back X-point position, X-point flux, or a multi-point LCFS, so
the shape rows in the table are open-loop observations and the evolution
panels~(b)--(d) sit systematically above either policy.  The AdvA-PPO gains
are not bought with larger actuation: its mean per-step $|\Delta V|$ is the
lowest of the three ($0.36\,\mathrm{V}$ against $0.56\,\mathrm{V}$ and
$0.45\,\mathrm{V}$).

\begin{figure}[htbp]
  \centering
  \includegraphics[width=\textwidth]{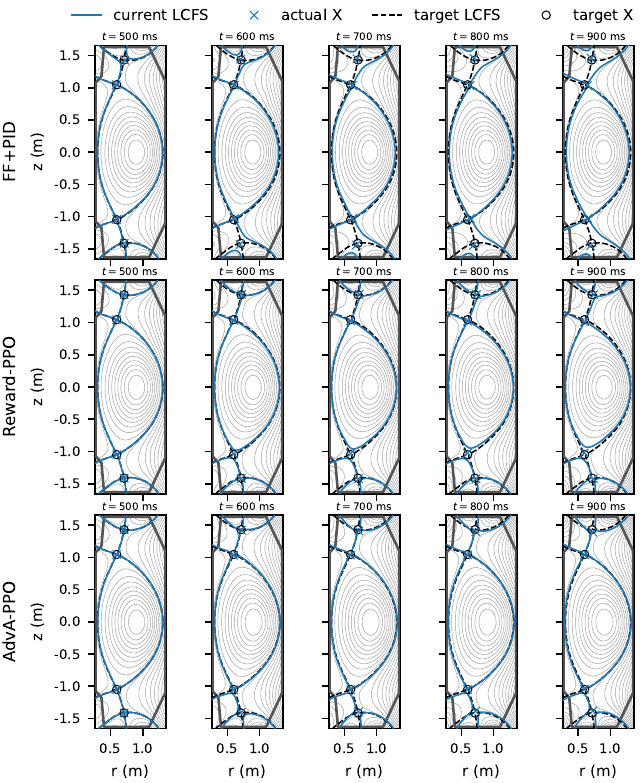}
  \caption{Flux snapshots on \#13906 at $t=500$--$900\,\mathrm{ms}$
  (shot clock; control from $t_0=500\,\mathrm{ms}$).
  Rows: FF+PID, Reward-PPO, AdvA-PPO.
  Blue solid: current LCFS; black dashed: target; open circles: target $X$;
  blue crosses: actual $X$.}
  \label{fig:flux-snapshots}
\end{figure}

Figure~\ref{fig:flux-snapshots} supplies the geometric reading of the same
episode.  Under AdvA-PPO the separatrix stays on the target boundary and the
X points remain close to their targets through the horizon, consistent with
the low primary distance and flux errors.  Reward-PPO keeps a four-X
topology but with a looser core LCFS match, in line with its intermediate
boundary and flux scores.  Under FF+PID the separatrix bulges outward and
the X points drift, as expected from a controller without shape closure.
Across all three methods the divertor-leg region remains only loosely
constrained: neither the observation nor the reward treats the legs
explicitly, so secondary-leg geometry is a shared limitation of the present
task rather than a failure unique to one controller.

\begin{table}[htbp]
  \centering
  \small
  \setlength{\tabcolsep}{5pt}
  \caption{Nominal tracking on I1 (\#13906, $500\,\mathrm{ms}$);
  metrics as in Sec.~\ref{sec:eval-metrics}.
  FF+PID X-point and LCFS rows are open-loop observations.}
  \label{tab:main-comparison}
  \begin{tabular}{@{}lccc@{}}
    \toprule
    Quantity & FF+PID & Reward-PPO & AdvA-PPO (ours) \\
    \midrule
    $I_p$ (A) & \underline{454} / 1488 & 735 / \underline{1310} & \textbf{260} / \textbf{624} \\
    Primary $X$ dist.\ (mm) & 39.0 / 58.3 & \underline{22.8} / \underline{39.2} & \textbf{6.3} / \textbf{9.5} \\
    Secondary $X$ dist.\ (mm) & 30.1 / \underline{38.3} & \textbf{19.0} / \textbf{31.8} & \underline{28.2} / 44.2 \\
    Mean $X$ flux ($10^{-4}\,\mathrm{Wb}$) & 77.38 / 117.02 & \underline{13.06} / \underline{18.89} & \textbf{0.66} / \textbf{1.93} \\
    LCFS ($\times 10^{-3}$) & 120.1 / 430.3 & \underline{14.7} / \underline{21.9} & \textbf{12.2} / \textbf{17.2} \\
    \midrule
    SmoothMax $\bar u$ $\uparrow$ & 0.093 & \underline{0.563} & \textbf{0.925} \\
    mean worst channel $\overline{r}$ $\uparrow$ & 0.001 & \underline{0.232} & \textbf{0.810} \\
    \midrule
    mean\,/\,max $|\Delta V|$ (V) & \underline{0.45} / 124 & 0.56 / \textbf{50} & \textbf{0.36} / \underline{81} \\
    \bottomrule
  \end{tabular}
\end{table}

\FloatBarrier

\subsection{Mechanism and ablation analysis}
\label{sec:mechanism-results}

\begin{table}[htbp]
  \centering
  \caption{Component ablation on I1 (\#13906, $500\,\mathrm{ms}$).
  Names as in Table~\ref{tab:controller-inventory};
  metrics as in Sec.~\ref{sec:eval-metrics}.}
  \label{tab:ablation}
  \scriptsize
  \setlength{\tabcolsep}{2.5pt}
  \begin{tabular}{@{}lcccccc@{}}
    \toprule
    Variant & $I_p$ (A) & mean $d_X$ (mm) & mean flux ($10^{-4}\,\mathrm{Wb}$) & LCFS ($\times 10^{-3}$) & $\bar u$ $\uparrow$ & $\overline{r}$ $\uparrow$ \\
    \midrule
    Reward-PPO & \underline{735} / 1310 & 20.9 / 35.5 & 13.06 / 18.89 & 14.7 / 21.9 & 0.563 & 0.232 \\
    Adv-mean & 1615 / 2149 & 34.2 / 45.8 & 1.77 / 2.70 & 8.7 / 14.9 & 0.830 & 0.665 \\
    Adv-SmoothMax & 1630 / 2545 & 17.6 / \underline{27.3} & 2.44 / 5.22 & 13.4 / 18.0 & 0.879 & 0.760 \\
    \quad + gate & 2779 / 3384 & 18.5 / 29.0 & 1.30 / 2.76 & 15.3 / 21.7 & \underline{0.899} & 0.760 \\
    \quad + resid $=$ \textbf{AdvA-PPO} & \textbf{260} / \textbf{624} & \underline{17.3} / \textbf{26.9} & \textbf{0.66} / \textbf{1.93} & 12.2 / 17.2 & \textbf{0.925} & \textbf{0.810} \\
    \quad + proj & 1601 / 1916 & 18.9 / 29.7 & \underline{1.27} / \underline{2.10} & 8.5 / \underline{11.3} & 0.894 & \underline{0.765} \\
    \quad + gate+resid & 1137 / 1445 & \textbf{16.8} / \underline{27.3} & 1.70 / 2.85 & \underline{7.5} / \textbf{8.8} & 0.894 & 0.748 \\
    \quad + gate+proj & 925 / \underline{1187} & 23.3 / 36.7 & 2.01 / 3.28 & \textbf{6.1} / \textbf{8.8} & 0.870 & 0.726 \\
    \bottomrule
  \end{tabular}
\end{table}

Table~\ref{tab:ablation} decomposes the nominal AdvA-PPO gain of
Sec.~\ref{sec:main-results} along the design sequence of
Table~\ref{tab:controller-inventory} (metrics in Sec.~\ref{sec:eval-metrics}).
Reward-PPO already holds $I_p$ reasonably ($735\,\mathrm{A}$ RMSE) but leaves
a large mean flux error ($13.1\times10^{-4}$), so the XPT is not flux-tight.
Moving credit assignment before scalarisation (Adv-mean, then Adv-SmoothMax)
cuts that flux error by about an order of magnitude, yet both raise $I_p$
RMSE above $1.6\,\mathrm{kA}$.
The two are not interchangeable: Adv-mean reaches a slightly lower flux
($1.77$ versus $2.44\times10^{-4}$) but a much worse mean $d_X$
($34.2$ versus $17.6\,\mathrm{mm}$), so the softmin temperature alone does not
select the operating point---geometry and current still trade.
On the Adv-SmoothMax anchor, $+\;$gate further lowers mean flux
($1.30\times10^{-4}$) while driving $I_p$ to $2.78\,\mathrm{kA}$ RMSE:
attenuating saturated channels sharpens the shape update and starves the
current loop.
Adding only $+\;$resid $=$ AdvA-PPO recovers both sides of that trade-off and
is the strongest balanced cell: relative to Adv-SmoothMax it reduces $I_p$
RMSE from $1.63$ to $0.26\,\mathrm{kA}$ and mean flux from $2.44$ to
$0.66\times10^{-4}$, with mean $d_X$ essentially unchanged
($17.3$ versus $17.6\,\mathrm{mm}$), and it also leads the shared scoring
axis ($\bar u=0.925$, $\overline{r}=0.810$).
That joint recovery is what links the family-level AdvA-PPO versus
Reward-PPO contrast of Sec.~\ref{sec:main-results} to a concrete component
choice.
The remaining proxy combinations do not overturn it:
$+\;$gate+proj records the best LCFS
($6.1\times10^{-3}$) but is weaker on $I_p$ ($925\,\mathrm{A}$) and mean
flux ($2.01\times10^{-4}$) than residual alone, so the boundary looks tidy
while the XPT flux condition and current loop do not;
$+\;$proj and $+\;$gate+resid likewise fail to match AdvA-PPO on $I_p$ or
mean flux.
Among gating, projection, and residual, the capped residual is therefore
retained: it is the most balanced correction and the one that keeps the
four-null flux lock that defines the XPT here.

\FloatBarrier

\subsection{Robustness evaluation}
\label{sec:robustness-results}

This subsection tests whether the same frozen controllers remain usable
under two deployment stresses.  First, diagnostic noise and observation
delay of the kind present in the experimental plant (R1--R3).  Second, a
change of the magnetic configuration at flattop hand-over: the controller
takes over near $t_0\approx 500\,\mathrm{ms}$ from a different parent
discharge while the \#13906 XPT geometric target stays fixed (I2--I4).
FF+PID, Reward-PPO, and AdvA-PPO are evaluated without retraining;
disturbance magnitudes, initials, and the clean-state scoring rule follow
Sec.~\ref{sec:eval-scenarios}.

\subsubsection{Zero-shot diagnostic noise and observation delay}
\label{sec:zero-shot-noise-delay}

\begin{figure}[htbp]
  \centering
  \includegraphics[width=\textwidth]{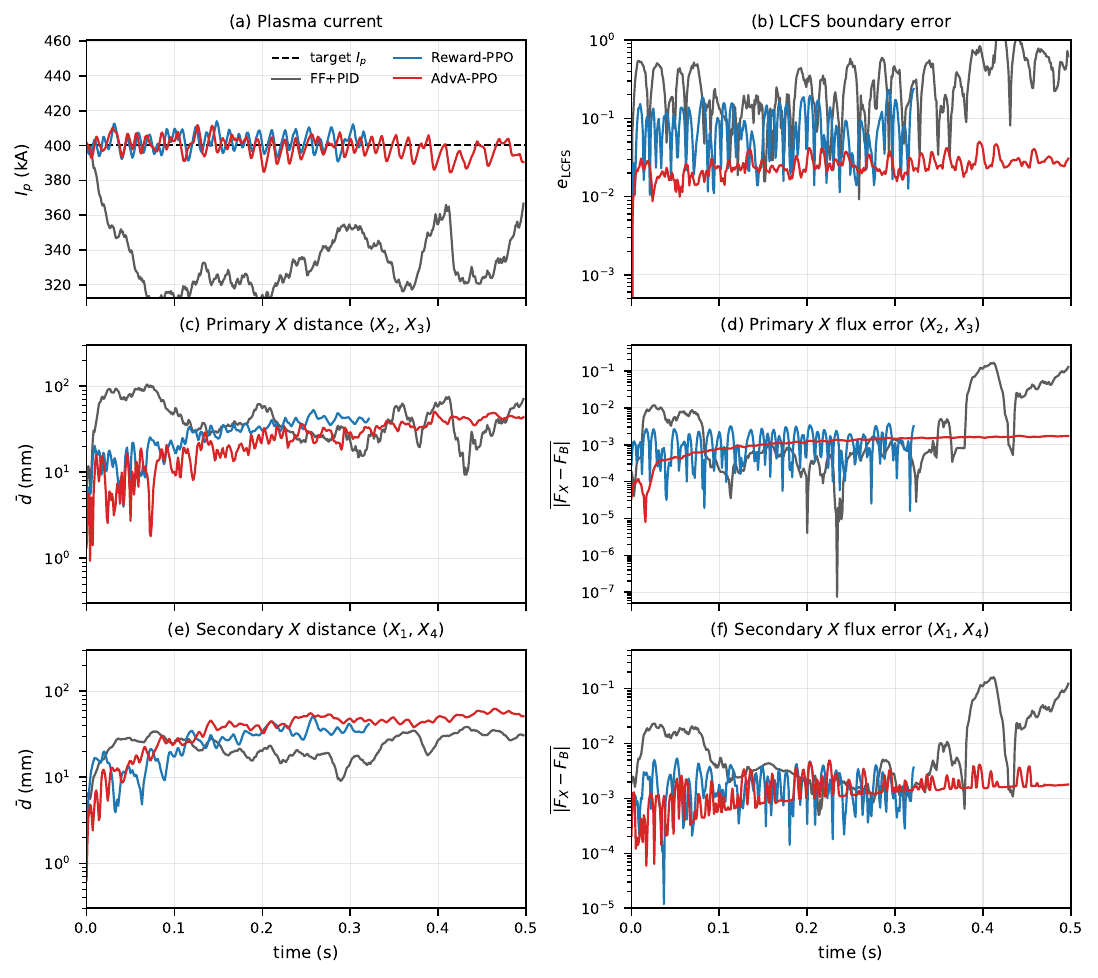}
  \caption{R3: delay + noise on I1 (\#13906).
  Panel layout as Figure~\ref{fig:main-comparison}.}
  \label{fig:robustness-r3}
\end{figure}

R1--R3 of Sec.~\ref{sec:eval-scenarios} stress the three controllers on I1
(\#13906).  Table~\ref{tab:robustness-noise-delay} reports survival and
RMSE over the survived horizon; Figure~\ref{fig:robustness-r3} shows the
channel-wise evolution under the combined delay+noise stress (R3).

\begin{table}[htbp]
  \centering
  \caption{Zero-shot noise/delay on I1 (\#13906); RMSE over the survived horizon (Sec.~\ref{sec:eval-metrics}).}
  \label{tab:robustness-noise-delay}
  \scriptsize
  \setlength{\tabcolsep}{2.5pt}
  \begin{tabular}{@{}llccccccc@{}}
    \toprule
    & Method & Surv.\ (ms) & $I_p$ (A) & Prim.\ $X$ (mm) & Prim.\ flux ($10^{-4}\,\mathrm{Wb}$) & Sec.\ $X$ (mm) & Sec.\ flux ($10^{-4}\,\mathrm{Wb}$) & LCFS ($\times 10^{-3}$) \\
    \midrule
    \multicolumn{9}{@{}l@{}}{\textit{R1: delay $1$--$3\,\mathrm{ms}$}} \\
    & FF+PID & 500 & \textbf{637} & 38.1 & \textbf{2.16} & \underline{29.6} & 149.81 & \underline{160.8} \\
    & Reward-PPO & 500 & 11997 & \textbf{23.0} & 28.38 & 41.1 & \underline{36.88} & 229.6 \\
    & AdvA-PPO & 500 & \underline{4521} & \underline{26.6} & \underline{13.04} & \textbf{24.8} & \textbf{17.85} & \textbf{18.8} \\
    \midrule
    \multicolumn{9}{@{}l@{}}{\textit{R2: full diagnostic noise}} \\
    & FF+PID & 500 & 74817 & 46.1 & 23.02 & \underline{22.3} & 55.54 & 222.0 \\
    & Reward-PPO & 500 & \textbf{2183} & \textbf{12.6} & \textbf{5.37} & \textbf{21.0} & \textbf{8.31} & \underline{53.0} \\
    & AdvA-PPO & 500 & \underline{2394} & \underline{26.4} & \underline{12.88} & 31.9 & \underline{12.60} & \textbf{19.3} \\
    \midrule
    \multicolumn{9}{@{}l@{}}{\textit{R3: delay + noise}} \\
    & FF+PID & \textbf{500} & 68752 & 48.6 & 368.72 & \textbf{24.9} & 356.89 & 422.3 \\
    & Reward-PPO & \underline{322} & \textbf{5222} & \underline{30.8} & \underline{15.17} & \underline{28.5} & \underline{21.39} & \underline{94.6} \\
    & AdvA-PPO & \textbf{500} & \underline{5521} & \textbf{29.0} & \textbf{12.79} & 41.7 & \textbf{17.82} & \textbf{26.3} \\
    \bottomrule
  \end{tabular}
\end{table}

The three conditions expose different failure modes rather than a single
winner.  On R1 (delay only) all three finish the horizon: FF+PID keeps the
best $I_p$ tracking ($637\,\mathrm{A}$ RMSE), consistent with a strong
feedforward backbone that is only weakly disturbed by a few milliseconds of
observation lag, while AdvA-PPO leads on LCFS and Reward-PPO suffers a large
$I_p$ RMSE under the per-step delay jitter.  On R2 (noise only) the ranking
reverses on the current channel: both learned controllers hold $I_p$ near
$2$--$3\,\mathrm{kA}$ RMSE, whereas FF+PID collapses to $>70\,\mathrm{kA}$.
That contrast is expected from the classical loop structure
(Sec.~\ref{sec:baseline-ff-pid}): derivative action on noisy $I_p$, coil-current,
and $R$/$Z$ measurements amplifies high-frequency sensor error into coil-voltage
corrections, so the same PID that is benign under clean delayed observations
becomes harmful under full diagnostic noise.  Within the learned pair on R2,
Reward-PPO is honestly the stronger cell---best $I_p$, primary/secondary
$X$ distance, and both flux channels---with AdvA-PPO retaining the best LCFS.
R3 combines both stresses and is the decisive joint test: Reward-PPO's R2
advantage does not carry over---it terminates at $322\,\mathrm{ms}$---while
AdvA-PPO and FF+PID complete the horizon (Figure~\ref{fig:robustness-r3}).
Among survivors AdvA-PPO keeps the lowest boundary and primary-flux errors;
FF+PID again survives but with the same noise-dominated current/flux/boundary
degradation seen on R2.  Taken together, the single-factor cells are
informative but not conclusive: FF+PID resists delay yet fails under noise,
and Reward-PPO wins the noise-only comparison yet breaks under the joint
load.  On the combined R3 protocol that matches concurrent plant delay and
diagnostic noise, AdvA-PPO is the strongest controller---the only learned
policy that both survives and retains a usable XPT shape.

\FloatBarrier

\subsubsection{Zero-shot cross-initialization}
\label{sec:zero-shot-cross-init}

\begin{figure}[htbp]
  \centering
  \includegraphics[width=\textwidth]{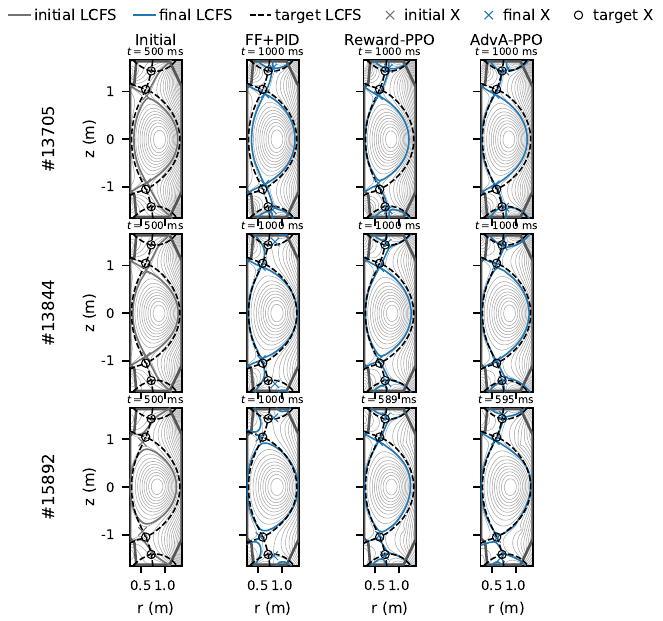}
  \caption{Zero-shot cross-init flux under frozen \#13906 XPT.
  Rows: \#13705, \#13844, \#15892.
  Columns: initial equilibrium and FF+PID / Reward-PPO / AdvA-PPO at
  episode end.
  Grey: initial LCFS/$X$; blue: final; black dashed/circles: target.}
  \label{fig:cross-init-flux}
\end{figure}

I2--I4 of Table~\ref{tab:eval-initials} hold the \#13906 XPT target,
observation map, and reward operators fixed, and change two quantities at
once: the FGE initial equilibrium and the $I_p$ reference (training-scale
$\sim\!400\,\mathrm{kA}$ to $\sim\!500\,\mathrm{kA}$).  That joint
shift---new shape or topology at hand-over plus a higher current
setpoint---is a hard zero-shot test for a policy trained on a single
initial.  Each controller keeps the checkpoint (or feedforward table)
already used on I1, with no per-initial redesign.  In particular, FF+PID
does not receive a newly computed feedforward for each withheld
discharge---that would amount to redesigning the classical controller and
is outside the present scope---so the classical baseline is the \#13906
plant feedforward and PID gains of Sec.~\ref{sec:baseline-ff-pid}, only
re-anchored at reset to remove the trivial CS/PF current offset (and with
the CS current guard removed).  We report survival and whether the frozen
XPT shape is reached (Figure~\ref{fig:cross-init-flux},
Table~\ref{tab:cross-init}).

\begin{table}[htbp]
  \centering
  \caption{Zero-shot cross-init to frozen \#13906 XPT (I2--I4); metrics as in Sec.~\ref{sec:eval-metrics} (full survival $=500\,\mathrm{ms}$). FF+PID feedforward is re-anchored as described in the text.}
  \label{tab:cross-init}
  \scriptsize
  \setlength{\tabcolsep}{2.2pt}
  \begin{tabular}{@{}llcccccccc@{}}
    \toprule
    Init.\ & Controller & Surv. & $I_p$ (kA) & mean $d_X$ (mm) & mean flux ($10^{-4}\,\mathrm{Wb}$) & LCFS ($\times 10^{-3}$) & $\bar u$ $\uparrow$ & $\overline{r}$ $\uparrow$ \\
    \midrule
    I2 \#13705 divertor & FF+PID & full & \textbf{9.4} / \textbf{42.3} & 125 / \textbf{178} & 1315 / \underline{5054} & 232 / 596 & 0.02 & \underline{0.00} \\
     & Reward-PPO & full & 32.8 / 79.3 & \underline{85} / \underline{179} & \textbf{889} / \textbf{5013} & \textbf{46} / \textbf{70} & \underline{0.07} & \underline{0.00} \\
     & AdvA-PPO & full & \underline{31.8} / \underline{75.6} & \textbf{69} / 179 & \underline{909} / 5187 & \underline{63} / \underline{129} & \textbf{0.11} & \textbf{0.01} \\
    \midrule
    I3 \#13844 divertor & FF+PID & full & \textbf{2.8} / \textbf{6.1} & 126 / \underline{174} & 1305 / \textbf{5009} & 97 / \underline{119} & 0.02 & 0.00 \\
     & Reward-PPO & full & 19.6 / \underline{56.8} & \underline{72} / \textbf{168} & \textbf{990} / \underline{5333} & \textbf{50} / \textbf{110} & \underline{0.10} & 0.00 \\
     & AdvA-PPO & full & \underline{17.6} / 57.3 & \textbf{60} / 176 & \underline{1061} / 5781 & \underline{79} / 241 & \textbf{0.14} & 0.00 \\
    \midrule
    I4 \#15892 limiter & FF+PID & \textbf{full} & \textbf{3.4} / \textbf{6.9} & \textbf{93} / \textbf{167} & \textbf{1168} / \underline{5082} & \underline{139} / \textbf{150} & \underline{0.02} & 0.00 \\
     & Reward-PPO & 90 & 31.4 / 43.2 & 103 / 173 & \underline{1697} / \textbf{5074} & \textbf{133} / 273 & \textbf{0.04} & 0.00 \\
     & AdvA-PPO & \underline{96} & \underline{20.7} / \underline{26.5} & \underline{99} / \underline{170} & 1714 / 5207 & 182 / \underline{272} & \textbf{0.04} & 0.00 \\
    \bottomrule
  \end{tabular}
\end{table}

Under this shift the classical baseline is the more transferable
stabiliser.  The same \#13906 FF+PID gains and feedforward, only
re-anchored, complete the horizon on every withheld initial and hold $I_p$
tightly ($2.8$--$9.4\,\mathrm{kA}$ RMSE against $17$--$33\,\mathrm{kA}$ for
the policies), spanning the training-scale $400\,\mathrm{kA}$ plant and the
$500\,\mathrm{kA}$ test setpoints.  That stability does not buy shape
precision: mean $d_X$ stays $\sim\!93$--$126\,\mathrm{mm}$, mean flux
errors remain large, and the limiter initial is held without converting to
the target XPT (Figure~\ref{fig:cross-init-flux}).

On the two divertor initials both policies survive and improve the geometric
channels relative to FF+PID (mean $d_X$ $60$--$85\,\mathrm{mm}$, lower LCFS
and flux RMSE), with AdvA-PPO leading on $d_X$ and $\bar u$ while Reward-PPO
keeps the lower LCFS error---so the advantage-level update is not uniformly
better off-distribution.  The current and boundary channels trade against
each other on \#13705, whose CS starts at $-15.7\,\mathrm{kA}$ and therefore
has about half the volt-second headroom of the other initials: FF+PID does
hold $I_p$ there, but only by drawing $54\,\mathrm{kA}$ of CS current against
$43$--$45\,\mathrm{kA}$ on the other two.

The limiter I4 (\#15892) fails both learned controllers: Reward-PPO
terminates at step~$90$ and AdvA-PPO at step~$96$, each with mean X-point
distance still $\sim\!100\,\mathrm{mm}$, while FF+PID survives without
converting to the target XPT (Table~\ref{tab:cross-init}).  The required
limiter-to-divertor topology change is absent from the AdvA-PPO training
distribution; the matched early stop of the Reward-PPO slot shows the
difficulty is shared.  FF+PID already travels across $I_p$ and initial as a
stabiliser, while the learned controller can still be adapted for the
missing conversion: that is the motivation for the multi-initialization
fine-tuning of Sec.~\ref{sec:finetune-robustness}.

\FloatBarrier

\subsubsection{Fine-tuning across magnetic initials}
\label{sec:finetune-robustness}

\begin{figure}[htbp]
  \centering
  \includegraphics[width=\textwidth]{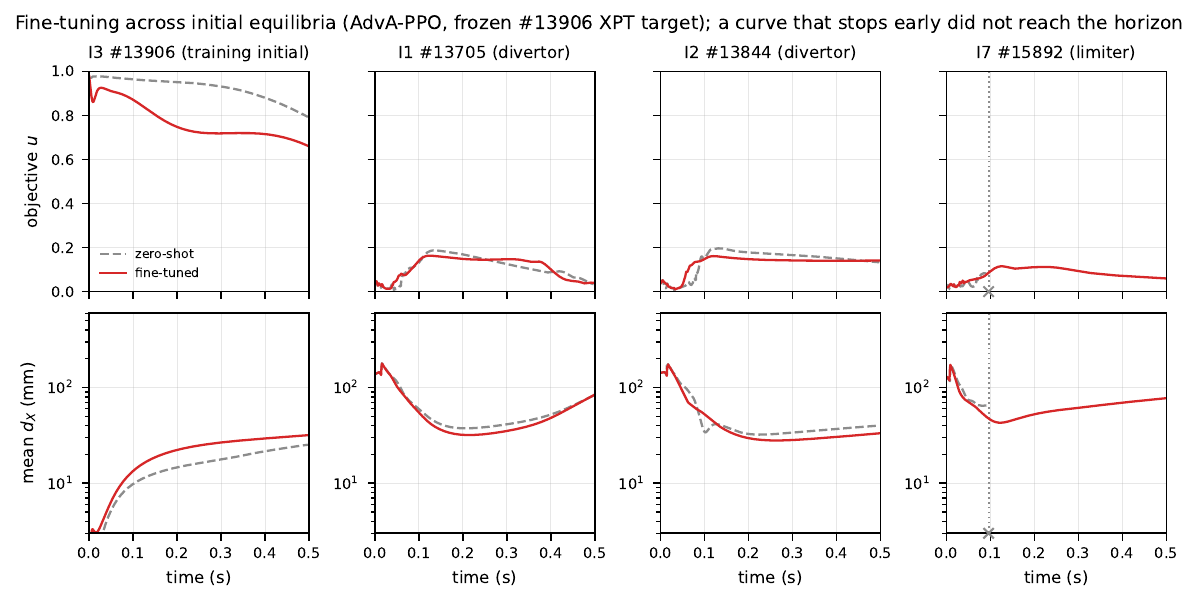}
  \caption{AdvA-PPO multi-initialization fine-tune ($100$ iterations,
  frozen \#13906 XPT).
  Grey dashed: zero-shot; red: fine-tuned.
  Top: $\bar u$; bottom: mean $d_X$.
  Early stop marked by a dotted vertical line.}
  \label{fig:finetune-cross-init}
\end{figure}

Zero-shot AdvA-PPO covers the two divertor initials but fails on the limiter
I4 (\#15892).  A short multi-initialization fine-tune of the frozen
checkpoint asks whether one adapted policy can span divertor and limiter
resets under the same frozen \#13906 XPT target.  The protocol widens only
the reset distribution: $100$ further iterations with a
Kullback--Leibler (KL) divergence-capped update
\cite{schulman2017ppo},
unchanged reward operators and observation map, and evaluation under the
harness of Sec.~\ref{sec:main-results}.  Because survival differs before and
after adaptation, Figure~\ref{fig:finetune-cross-init} and
Table~\ref{tab:finetune} summarise the before/after comparison; every error
and objective entry in the table is taken on the common window of its
before/after pair, with survival reported separately.

\begin{table}[htbp]
  \centering
  \caption{AdvA-PPO before/after multi-initialization fine-tuning
  ($100$ iterations, frozen \#13906 XPT).
  Errors and $\bar u$/$\overline{r}$ on the common window of each pair;
  survival separate.
  On \#15892 the common window is the $96$ zero-shot steps; over its own
  $500\,\mathrm{ms}$ the fine-tuned run reaches $\bar u=0.08$, $I_p$ RMSE
  $20.8\,\mathrm{kA}$, mean $d_X=74\,\mathrm{mm}$.}
  \label{tab:finetune}
  \scriptsize
  \setlength{\tabcolsep}{3pt}
  \begin{tabular}{@{}llcccccc@{}}
    \toprule
    Init.\ & Controller & Surv. & $I_p$ (kA) & mean $d_X$ (mm) & LCFS ($\times 10^{-3}$) & $\bar u$ $\uparrow$ & $\overline{r}$ $\uparrow$ \\
    \midrule
    I1 \#13906 (XPT / training) & zero-shot & full & \textbf{0.3} & \textbf{17} & \textbf{12} & \textbf{0.92} & \textbf{0.81} \\
     & fine-tuned & full & \underline{0.7} & \underline{24} & \underline{17} & \underline{0.77} & \underline{0.58} \\
    I2 \#13705 (divertor) & zero-shot & full & \underline{31.8} & \underline{69} & \textbf{63} & \textbf{0.11} & \textbf{0.01} \\
     & fine-tuned & full & \textbf{26.2} & \textbf{67} & \underline{64} & \textbf{0.11} & \underline{0.00} \\
    I3 \#13844 (divertor) & zero-shot & full & \underline{17.6} & \underline{60} & \underline{79} & \textbf{0.14} & \textbf{0.00} \\
     & fine-tuned & full & \textbf{13.7} & \textbf{58} & \textbf{66} & \underline{0.13} & \textbf{0.00} \\
    I4 \#15892 (limiter) & zero-shot & \underline{96} & \textbf{20.7} & \underline{99} & \underline{182} & \underline{0.04} & \textbf{0.00} \\
     & fine-tuned & \textbf{full} & \underline{25.0} & \textbf{92} & \textbf{88} & \textbf{0.05} & \textbf{0.00} \\
    \bottomrule
  \end{tabular}
\end{table}

The fine-tuned policy covers all four initials
(Figure~\ref{fig:finetune-cross-init}).  On \#15892 survival rises
from $96$ steps to the full horizon; over the shared $96$ steps the
objective is essentially unchanged ($0.04$ against $0.05$) while the
boundary error halves ($182$ to $88\times10^{-3}$), so the budget buys the
remaining horizon and a cleaner boundary rather than a uniformly stronger
tracker.  On the two divertor initials, already full-horizon zero-shot, the
objective stays flat ($0.11$ to $0.11$ on \#13705, $0.14$ to $0.13$ on
\#13844), with a modest $I_p$ gain ($31.8$ to $26.2$ and $17.6$ to
$13.7\,\mathrm{kA}$ RMSE) and geometric channels essentially held.  The
inevitable compromise appears on the training initial \#13906: $\bar u$
drops from $0.92$ to $0.77$ and the mean worst channel from $0.81$ to
$0.58$.  Multi-initialization coverage and nominal specialisation are
therefore not free jointly under this short budget---the same weights that
span divertor and limiter give up part of the \#13906 solution.

Each cell is a single trajectory, so survival figures are attempted
rollouts rather than success rates.  Fine-tuning under the noise and delay
conditions of Sec.~\ref{sec:zero-shot-noise-delay} is left for future work.

\section{Conclusion}
We established an AI-enabled multi-objective framework for EXL-50U XPT
magnetic control, formulated as reconstructed-null feedback through shared
poloidal-field actuators.  Its ten physical objectives jointly represent
$I_p$, the LCFS, four X-point positions, and four X-point flux constraints.
The FGE environment reproduces the experimental \#13906 equilibrium at
centimetre-scale boundary and null accuracy, providing an
experiment-calibrated free-boundary test bed for closed-loop evaluation.

\textbf{AdvA} addresses multi-objective temporal credit assignment by retaining
one value head and one GAE per channel on a shared critic backbone, applying
worst-objective-aware SmoothMax only
after advantage estimation, and adding a norm-capped component residual to
the aggregate policy update.  On the nominal $500\,\mathrm{ms}$ rollout,
AdvA-PPO raises the shared evaluation score $\bar u$ from 0.56 to 0.93 and the
mean worst-channel score from 0.23 to 0.81 relative to Reward-PPO, while
reducing mean X-point flux RMSE by about $20\times$ and using the lowest mean
per-step voltage change.  Among the evaluated corrections, the capped
residual gives the most balanced recovery of $I_p$ and X-point flux accuracy;
the result favours a controlled residual over stacking every available
multi-objective correction.

The broader evaluation reveals complementary, regime-dependent strengths.
Under combined measurement noise and observation delay, AdvA-PPO is the only
learned controller that completes the horizon while retaining a usable XPT
shape, although Reward-PPO is stronger under noise alone and the
experiment-derived FF+PID baseline tracks $I_p$ well under delay.  Across
withheld initial equilibria, the same FF+PID controller is the more
transferable stabiliser but remains shape-imprecise and does not produce the
limiter-to-XPT conversion; the learned controllers improve geometric
regulation on the diverted initials but both fail that conversion zero-shot.
Multi-initial adaptation extends AdvA-PPO to all four initials and exposes a
coverage--specialisation trade-off.  Together, these results establish the
capability and current boundaries of reconstructed-null XPT feedback in an
experiment-calibrated magnetic simulation; PTEFIT-in-the-loop and closed-loop
machine performance remain to be tested experimentally.

\section{Future outlook}
The immediate next step is staged on-machine validation on EXL-50U.  The
near-term campaign will first quantify the accuracy, latency, and stability of
PTEFIT secondary-null position and flux estimates, and then evaluate the
present RL controller with PTEFIT in the loop.  Conservative actuator limits,
online monitoring, and a deterministic FF+PID fallback will allow the effects
of real-time reconstruction and learned control to be assessed separately
before extending the tests to disturbances and broader operating conditions.

Algorithmic development will target robustness and generalisation beyond the
single-initial training regime.  Multi-initial and uncertainty-aware training
should span plasma-current targets, measurement uncertainties, actuator
errors, and FGE calibration residuals.  Domain randomisation, constrained
policy optimisation, and stage-dependent objectives provide complementary
routes to wider coverage while limiting the loss of nominal precision exposed
by the present fine-tuning study.

The control task should also expand from a fixed XPT target to broader
divertor configurations and controlled topology transitions.  Strike-point
and divertor-leg targets, richer LCFS descriptors, and explicit coil-current,
voltage, and slew constraints are natural next objectives.  Coordinating these
tests with auxiliary-heating experiments will probe magnetic-configuration
control under more relevant plasma conditions and prepare later integration
with radiation, detachment, and target-heat-flux control as suitable models
and diagnostics become available.  Throughout this progression, hard safety
limits, interlocks, fallback control, and staged validation must make machine
protection part of the controller design rather than an external safeguard.

\clearpage  % flush any remaining floats before the bibliography
\bibliographystyle{unsrt}
\bibliography{references}

@article{kuang2020sparc,
  author  = {Kuang, A. Q. and others},
  title   = {Divertor heat flux challenge and mitigation in SPARC},
  journal = {Journal of Plasma Physics},
  volume  = {86},
  year    = {2020},
  doi     = {10.1017/S0022377820001117}
}

@article{labombard2015,
  author  = {LaBombard, B. and others},
  title   = {ADX: a high field, high power density, advanced divertor and RF tokamak},
  journal = {Nuclear Fusion},
  volume  = {55},
  pages   = {053020},
  year    = {2015},
  doi     = {10.1088/0029-5515/55/5/053020}
}

@article{umansky2017assessment,
  author  = {Umansky, M. V. and others},
  title   = {Assessment of X-point target divertor configuration for power handling and detachment front control},
  journal = {Nuclear Materials and Energy},
  year    = {2017},
  doi     = {10.1016/j.nme.2017.03.015}
}

@article{theiler2017,
  author  = {Theiler, C. and others},
  title   = {Results from recent detachment experiments in alternative divertor configurations on TCV},
  journal = {Nuclear Fusion},
  volume  = {57},
  pages   = {072008},
  year    = {2017},
  doi     = {10.1088/1741-4326/aa5fb7}
}

@article{raj2022,
  author  = {Raj, H. and others},
  title   = {Improved heat and particle flux mitigation in high core confinement, baffled, alternate divertor configurations in the TCV tokamak},
  journal = {Nuclear Fusion},
  year    = {2022},
  doi     = {10.1088/1741-4326/ac94e5}
}

@article{lee2025,
  author  = {Lee, K. and others},
  title   = {X-Point Target Radiator Regime in Tokamak Divertor Plasmas},
  journal = {Physical Review Letters},
  volume  = {134},
  pages   = {185102},
  year    = {2025},
  doi     = {10.1103/PhysRevLett.134.185102}
}

@article{anand2024,
  author  = {Anand, H. and others},
  title   = {Real-time plasma equilibrium reconstruction and shape control for the MAST Upgrade tokamak},
  journal = {Nuclear Fusion},
  volume  = {64},
  pages   = {086051},
  year    = {2024},
  doi     = {10.1088/1741-4326/ad5c80}
}

@article{kolemen2018diiid,
  author  = {Kolemen, E. and others},
  title   = {Initial development of the DIII-D snowflake divertor control},
  journal = {Nuclear Fusion},
  volume  = {58},
  pages   = {066007},
  year    = {2018},
  doi     = {10.1088/1741-4326/aab0d3}
}

@article{soukhanovskii2016,
  author  = {Soukhanovskii, V. A. and others},
  title   = {Snowflake Divertor Experiments in the DIII-D, NSTX, and NSTX-U Tokamaks Aimed at the Development of the Divertor Power Exhaust Solution},
  journal = {IEEE Transactions on Plasma Science},
  volume  = {44},
  number  = {12},
  pages   = {3445--3455},
  year    = {2016},
  doi     = {10.1109/TPS.2016.2625325}
}

@misc{lonigro2026,
  author        = {Lonigro, N. and others},
  title         = {Initial observations in X-point target divertor discharges on MAST-U},
  year          = {2026},
  eprint        = {2601.21840},
  archivePrefix = {arXiv}
}

@article{gu2025ehl2,
  author  = {Gu, Xiang and others},
  title   = {Poloidal field system and advanced divertor equilibrium configuration design of the EHL-2 spherical torus},
  journal = {Plasma Science and Technology},
  volume  = {27},
  pages   = {024011},
  year    = {2025},
  doi     = {10.1088/2058-6272/adae72}
}

@article{wang2025ehl2,
  author  = {Wang, Fuqiong and others},
  title   = {Divertor heat flux challenge and mitigation in the EHL-2 spherical torus},
  journal = {Plasma Science and Technology},
  volume  = {27},
  pages   = {024009},
  year    = {2025},
  doi     = {10.1088/2058-6272/adadb8}
}

@article{shi2026exl,
  author  = {Shi, Yuejiang and others},
  title   = {Overview of EXL-50U experiments: addressing key physics issues for future spherical torus reactors},
  journal = {Nuclear Fusion},
  volume  = {66},
  pages   = {116009},
  year    = {2026},
  doi     = {10.1088/1741-4326/ae738d}
}

@article{moret2015liuqe,
  author  = {Moret, Jean-Marc and Duval, B. P. and Le, H. B. and Coda, S. and Felici, F. and Reimerdes, H.},
  title   = {Tokamak equilibrium reconstruction code LIUQE and its real time implementation},
  journal = {Fusion Engineering and Design},
  volume  = {91},
  pages   = {1--15},
  year    = {2015},
  doi     = {10.1016/j.fusengdes.2014.09.019}
}

@article{heiss2026fge,
  author  = {Hei{\ss}, Cosmas and Merle, Antoine and Carpanese, Francesco and Felici, Federico and Donner, Craig and Marchioni, Stefano and Mari, Alessandro and Sauter, Olivier},
  title   = {FGE: a fast free-boundary Grad--Shafranov evolutive solver},
  journal = {Plasma Physics and Controlled Fusion},
  volume  = {68},
  pages   = {045031},
  year    = {2026},
  doi     = {10.1088/1361-6587/ae56b7}
}

@article{degrave2022,
  author  = {Degrave, Jonas and others},
  title   = {Magnetic control of tokamak plasmas through deep reinforcement learning},
  journal = {Nature},
  volume  = {602},
  pages   = {414--419},
  year    = {2022},
  doi     = {10.1038/s41586-021-04301-9}
}

@article{subbotin2026,
  author  = {Subbotin, Artem and others},
  title   = {Demonstration of reconstruction-free static magnetic control of DIII-D plasma with deep reinforcement learning},
  journal = {Nuclear Fusion},
  year    = {2026},
  doi     = {10.1088/1741-4326/ae34c6}
}

@article{gorno2024,
  author  = {Gorno, S. and F{\'e}vrier, O. and Theiler, C. and Ewalds, T. and Felici, F. and Lunt, T. and Merle, A. and Degrave, J. and Duval, B. P. and Lee, K. and Reimerdes, H. and Tracey, B. and Wischmeier, M. and W{\"u}thrich, C.},
  title   = {X-point radiator and power exhaust control in configurations with multiple X-points in TCV},
  journal = {Physics of Plasmas},
  volume  = {31},
  pages   = {072504},
  year    = {2024},
  doi     = {10.1063/5.0201401}
}

@article{tracey2024,
  author  = {Tracey, B. D. and Michi, A. and Chervonyi, Y. and Davies, I. and Paduraru, C. and Lazic, N. and Felici, F. and Ewalds, T. and Donner, C. and Galperti, C. and Buchli, J. and Neunert, M. and Huber, A. and Evens, J. and Kurylowicz, P. and Mankowitz, D. J. and Riedmiller, M.},
  title   = {Towards practical reinforcement learning for tokamak magnetic control},
  journal = {Fusion Engineering and Design},
  volume  = {200},
  pages   = {114161},
  year    = {2024},
  doi     = {10.1016/j.fusengdes.2024.114161}
}

@article{kerboua2024west,
  author  = {Kerboua-Benlarbi, S. and Nouailletas, R. and Faugeras, B. and Nardon, E. and Moreau, P.},
  title   = {Magnetic control of {WEST} plasmas through deep reinforcement learning},
  journal = {IEEE Transactions on Plasma Science},
  year    = {2024},
  doi     = {10.1109/TPS.2024.3377811}
}

@misc{mele2025mpc,
  author        = {Mele, A. and Topalova, M. A. and Galperti, C. and Coda, S.},
  title         = {First experimental demonstration of plasma shape control in a tokamak through Model Predictive Control},
  year          = {2025},
  eprint        = {2506.20096},
  archivePrefix = {arXiv}
}

@misc{winkel2026,
  author        = {Winkel, M. and Verhaegh, K. and Kool, B. and Lee, K. and Carpita, M. and Perek, A. and Morgan, R. and Derks, G. and F{\'e}vrier, O. and Theiler, C. and Brida, D. and van Berkel, M.},
  title         = {Detachment dynamics and disturbance rejection in the {TCV} X-Point Target divertor},
  year          = {2026},
  eprint        = {2606.23432},
  archivePrefix = {arXiv}
}

@misc{schulman2017ppo,
  author        = {Schulman, John and Wolski, Filip and Dhariwal, Prafulla and Radford, Alec and Klimov, Oleg},
  title         = {Proximal Policy Optimization Algorithms},
  year          = {2017},
  eprint        = {1707.06347},
  archivePrefix = {arXiv}
}

@misc{schulman2016gae,
  author        = {Schulman, John and Moritz, Philipp and Levine, Sergey and Jordan, Michael and Abbeel, Pieter},
  title         = {High-Dimensional Continuous Control Using Generalized Advantage Estimation},
  year          = {2016},
  eprint        = {1506.02438},
  archivePrefix = {arXiv}
}

@article{mnih2015dqn,
  author  = {Mnih, Volodymyr and others},
  title   = {Human-level control through deep reinforcement learning},
  journal = {Nature},
  volume  = {518},
  pages   = {529--533},
  year    = {2015},
  doi     = {10.1038/nature14236}
}

@misc{zheng2026ptefit,
  author        = {Zheng, G. H. and Liu, S. F. and Gu, X. and Zhang, Y. P. and Li, J. and Liu, Y. and Lun, X. C. and Xing, L. and Chen, J. G. and Chen, Z. Y. and Yu, Y. and Guo, D. and Yang, Z. Y. and Xie, H. S. and Song, X. M. and Shi, Y. J. and {EXL-50U Team}},
  title         = {A Novel Numerical Algorithms Optimization Method with Machine Learning Frameworks: Application on Real-time Plasmas Equilibrium Reconstruction in {EXL}-50{U} Spherical Torus},
  year          = {2026},
  eprint        = {2601.12378},
  archivePrefix = {arXiv}
}

@inproceedings{yu2020pcgrad,
  author    = {Yu, Tianhe and Kumar, Saurabh and Gupta, Abhishek and Levine, Sergey and Hausman, Karol and Finn, Chelsea},
  title     = {Gradient surgery for multi-task learning},
  booktitle = {Advances in Neural Information Processing Systems},
  volume    = {33},
  year      = {2020}
}

@misc{munn2025gcr,
  author        = {Munn, Humphrey and Tidd, Brendan and Boehm, Peter and Gallagher, Marcus and Howard, David},
  title         = {Scalable Multi-Objective Robot Reinforcement Learning through Gradient Conflict Resolution},
  year          = {2025},
  eprint        = {2509.14816},
  archivePrefix = {arXiv}
}

@misc{ambadkar2026d3po,
  author        = {Ambadkar, Tanmay and Panda, Sourav and Kale, Shreyash and Dodge, Jonathan and Verma, Abhinav},
  title         = {Preference Conditioned Multi-Objective Reinforcement Learning: Decomposed, Diversity-Driven Policy Optimization},
  year          = {2026},
  eprint        = {2602.07764},
  archivePrefix = {arXiv}
}

@article{song2019hl2m,
  author  = {Song, X. M. and Li, J. X. and Leuer, J. A. and Zhang, J. H. and Song, X.},
  title   = {First plasma scenario development for {HL-2M}},
  journal = {Fusion Engineering and Design},
  volume  = {147},
  pages   = {111254},
  year    = {2019},
  doi     = {10.1016/j.fusengdes.2019.111254}
}

\end{document}